\documentclass{aa}
\usepackage{graphicx}
\usepackage{amssymb}
\usepackage{url}
\def\ssim{\setbox0=\hbox{$\sim$}%
\setbox1=\hbox{$<$}\dimen0=\ht1%
\advance\dimen0by-1.2pt\,\lower.6\dimen0%
\copy0\kern-\wd0\raise.4\dimen0\copy1 \,}

\def\gsim{\setbox0=\hbox{$\sim$}%
\setbox1=\hbox{$>$}\dimen0=\ht1%
\advance\dimen0by-1.2pt\,\lower.6\dimen0%
\copy0\kern-\wd0\raise.4\dimen0\copy1\,}

\def\lambdab{\lambda\mkern-9mu\lower1.2pt\hbox{$\mathchar'26$}}%

\begin{document}
   \title{The early star generations: the dominant effect of rotation on the CNO yields}

\authorrunning{G. Meynet, S. Ekstr\"om, \& A. Maeder}
\titlerunning{The early star generations}  

 \author{G. Meynet,
           S. Ekstr\"om
            \&
	     A. Maeder}

   \offprints{G. Meynet\\ email: Georges.Meynet@obs.unige.ch}

   \institute{Geneva Observatory CH -- 1290 Sauverny, Switzerland
             }

   \date{Received / Accepted }

\abstract{}{We examine the role of rotation on the evolution and chemical yields of very metal--poor stars.} 
 {The models include the same physics, which was applied successfully at the solar $Z$ and for the SMC, in particular, 
 shear diffusion, meridional circulation, horizontal turbulence, and rotationally enhanced mass loss.} 
 {Models of very low $Z$ experience a much stronger internal mixing in all phases than at solar $Z$. Also, 
rotating models at very low $Z$, contrary to the usual considerations, show a large mass loss, 
which mainly results  from the efficient mixing of the
products of the 3$\alpha$ reaction into the H--burning shell. This mixing allows convective dredge--up to enrich the stellar surface
in heavy elements during the red supergiant phase, which in turn favours a large loss of mass by stellar winds, 
especially as rotation also 
increases the duration of this phase.
On the whole,
the low $Z$ stars may lose about half of their mass. 
Massive stars initially rotating at half of their critical
velocity are likely to avoid the pair--instability supernova.
The chemical composition of the rotationally enhanced winds of very low $Z$ stars show 
large CNO enhancements  by factors of $10^3$ to $10^7$, together with large excesses of $^{13}$C
and $^{17}$O  and moderate amounts of Na and Al. The excesses of primary N are particularly striking.
When these ejecta from the rotationally
enhanced winds are diluted with the supernova ejecta from the corresponding CO cores, we find [C/Fe], [N/Fe],[O/Fe]
abundance ratios that are very similar to those observed in the C--rich, extremely metal--poor stars (CEMP).
We show that rotating AGB stars and rotating massive stars have about the same effects on the CNO enhancements.
Abundances of s-process elements and the $^{12}$C/$^{13}$C ratio could help us to distinguish between contributions
from AGB and massive stars.}{}
%{On the whole, we emphasize the dominant effects of rotation for the chemical yields.
%of extremely metal poor stars. Rotation has much more effects than changes of $Z$.}
\keywords {Stars: evolution -- Stars: rotation -- Stars: abundances -- Stars: mass loss}

 \maketitle
%
%________________________________________________________________

\section{Introduction}

Interest in the evolution of extremely metal--poor stars
has been stimulated recently by at least two types
of observing programmes.
%How did evolve the first massive stars in the Universe~? 
%What was their ionizing power~? How did they enrich their
%surroundings in newly synthesized elements~?
%For many years these questions were considered as academic ones.
%Indeed the first stars
%have since a long time disappeared in our near environment, leaving us no chance to directly
%compare models with observations. Moreover the metallicity
%increase of the interstellar medium is a rapid process
%making the period during which extremely metal poor stars govern  the chemical evolution of galaxies
%very short. Typically according to Audouze \& Silk (\cite{Au95}), 
%in the frame of the simple closed box model for the chemical evolution of
%our Galaxy,
%a metallicity equal to $\sim 2 \times 10^{-4}$ of the solar value ({\it i.e.}, [Fe/H]$\sim$ -3.7)
%is reached after a period of $\sim 2 \times 10^{6}$ yr, corresponding
%to the lifetime of only one generation of massive stars\footnote{Such a rough estimate,
%which neglects the mixing timescale and the infall, underestimates
%the timescale for the growth of the metallicity in the interstellar medium of our Galaxy. 
%On the other hand it very well illustrates the fact that massive stars are very efficient
%sources of heavy elements.}.
First, the detection of very far galaxies at redshifts well beyond 6 (see e.g. Pell\'o et al.~\cite{Pe05})
opens the way to detection of galaxies whose colours 
will be dominated by extremely metal--poor stars (Schaerer~\cite{Sc02}; \cite{Sc03}). 
Second, as a complement to the observation of the deep Universe, 
the detection 
of nearby, very metal--poor halo stars provides very interesting clues
to the early chemical evolution of our Galaxy
(Beers et al.~\cite{Be92}; Beers~\cite{Be99}; Christlieb et al. \cite{christ04}; Bessell et al.
\cite{bessel04}; Cayrel et al. \cite{cayr04}; Spite et al.\cite{Sp05}). 
These works have shown
the following
very interesting results:
\begin{itemize}
\item {\it The measured abundances of many elements at very low metallicity present
a very small scatter (Cayrel et al.~\cite{cayr04}).}
At first sight this appears difficult to understand. Indeed,
in the early Universe, stars are 
believed to form from the ejecta of
a small number of supernovae (may be only one). 
For instance the Argast et al. models (\cite{Ar00}; \cite{Ar02}) predict that for
[Fe/H] $< -3$, type II supernovae enrich the interstellar medium only locally.
Since the chemical composition of supernova
ejecta may differ a lot from case to case, large scatter 
of the abundances is expected
at very low metallicity. 
For most of the elements, however, this strong scatter is not observed,
at least down to a metallicity of [Fe/H]$\sim$ -4.0
(Cayrel et al. \cite{cayr04}).
This might indicate that, already at this low metallicity, stars are formed from 
a well--mixed reservoir composed of ejecta from stars of different initial masses.
\item {\it These observations also show that there is no sign of enrichments
by pair--instability supernovae,
at least down to a metallicity of [Fe/H] equal to -4.} Let us recall that these supernovae have 
very massive stars as progenitors, with initial masses between approximately 140 and 260 M$_\odot$ 
(Barkat et al. \cite{Ba67}; Bond et al. \cite{Bo84}; Heger \& Woosley~\cite{HW02}). 
Such massive stars are believed to form only in very metal--poor environments.
At the end of their lifetime,
they are completely destroyed when they explode as a pair--instability supernova. In this way
they may strongly contribute to the enrichment of the primordial interstellar medium
in heavy elements.
The composition of the ejecta of pair--instability supernovae is characterised by
a well marked odd-even effect and a strong zinc deficiency. These two features are not observed
in the abundance pattern of very metal--poor halo stars.
Does this mean that at [Fe/H] equal to -4,
pair--instability supernovae
no longer dominate the chemical evolution of galaxies or that such stars are not formed~?
If formed, could these stars skip the pair instability or have different nucleosynthetic outputs
from those currently predicted by theoretical models~?
\item {\it The N/O ratios observed at the surface of halo stars by Israelian et al.~(\cite{Is04}) and
Spite et al. (\cite{Sp05})
indicate that important amounts of primary nitrogen should be produced by 
very metal--poor massive stars
(Chiappini et al.~\cite{Ch05}).} The physical conditions for such important
productions of primary nitrogen by very metal--poor massive stars remain to be found.
\item {\it If most stars at a given [Fe/H] present a great
homogeneity in composition, a small group, comprising about 20 - 25\% of the stars
with [Fe/H] below -2.5, show very large enrichments in carbon.}
These stars are known as C-rich
extremely metal--poor (CEMP) stars. The observed [C/Fe] ratios
are between
$\sim$2 and 4, showing a large scatter.
Other elements, such as nitrogen and oxygen (at least in the few cases
where the abundance of this element could be measured), are also highly  
enhanced. Interestingly, 
the two most metal--poor stars known up to now, the Christlieb star
or HE 0107-5240, a halo giant with [Fe/H]=-5.3,
and the subgiant or main-sequence star HE 1327-2326 with [Fe/H]=-5.4 (Frebel et al.~\cite{Fr05})
belong in this category.
To explain such high and scattered CNO abundances, 
obviously a special process has to be invoked
(see Sect.~6 below). 
\end{itemize}
The results outlined above 
clearly indicate that new scenarios for the
formation and evolution of massive stars at very low $Z$
need to be explored.

Among the physical ingredients that could open new evolutionary paths
in very metal--poor environments,
rotation certainly appears a very interesting possibility.  
First, for metallicities $Z$ between 0.004 and 0.040, 
the inclusion of rotation improves the agreement between the models and 
observations in many respects by allowing us to reproduce the observed
surface abundances (Heger \& Langer \cite{He00}; Meynet \& Maeder \cite{MMV}), 
the ratio of blue--to--red supergiants at low metallicity (Maeder \& Meynet \cite{MMVII}),
the variation with the metallicity of the WR/O ratios 
and of the numbers of type Ibc to type II supernovae 
(Meynet \& Maeder \cite{MMX}; \cite{MMXI}). Most
likely, stars are also rotating at very low metallicity, and one can 
hope that the same physical model assumptions that
improve the physical description of stars at $Z \ge$ 0.004 
would also apply to the very low
metallicity domain. 

Second, if the effects of rotation are already 
quite significant at high metallicity,
one expects that they are even more important at lower metallicity. 
For instance, it was shown in previous
works that 
the chemical mixing becomes more efficient for lower
metallicity  for a given  initial mass and velocity 
(Maeder \& Meynet \cite{MMVII}; Meynet \& Maeder \cite{MMVIII}). 
This comes from the fact that
the gradients of $\Omega$ are much steeper in the lower metallicity
models, so they trigger more efficient shear mixing.
The gradients are steeper because 
less angular momentum is transported outwards by the
meridional currents, whose velocity
scales as the inverse of the density in the outer layers
(see the Gratton-\"Opick term in the expression for the meridional velocity in Maeder \& Zahn~\cite{mz98}).
 
Third, rotation can induce mass loss in two ways.
The first way, paradoxically, is linked to the fact
that very metal--poor  
stars are believed to lose little mass by radiatively driven stellar winds.
Indeed, in the radiatively driven wind theory, 
the mass loss scales with the metallicity of the outer
stellar layers as $(Z/{\rm Z}_\odot)^{\alpha}$ with $\alpha$ between 0.5 and 0.8
(Kudritzki et al.~\cite{Kud87}; Vink \& al. \cite{vink01}). Thus lowering the metallicity
by a factor 200 000 (as would be the 
case for obtaining the metallicity of the Christlieb star)
would thus lower the mass loss rates by a factor 450,
or even by a greater factor if the metallicity dependence becomes stronger
at lower $Z$, as suggested by Kudritzki~(\cite{Ku02}).
Now since metal--poor stars lose little mass, 
they also lose little angular momentum (if rotating),
so they have a greater chance of
reaching the break-up limit during the Main Sequence phase 
(see for instance Fig.~9 in Meynet \& Maeder~\cite{MMVIII}).
At break-up, the outer stellar layers become unbound and 
are ejected whatever their metallicity.  
The break-up is reached more easily when we take into
account that
massive rotating stars have polar winds as shown by Owocki et al.~(\cite{Ow96}) 
and Maeder~(\cite{MaIV}). When most of the mass is lost
along the rotational axis, little angular momentum is lost.

Another way for rotation to trigger enhancements of the mass loss
comes from the mixing induced by rotation. In general, rotational mixing
favours the evolution into the red supergiant stage (see Maeder \& Meynet \cite{MMVII}), where mass loss is higher. It also
enhances the metallicity of the surface of the star and, in this way, boosts the radiatively driven
stellar winds (see below).

Thus there are very good reasons for exploring the effects of rotation at very
low metallicity, which we have attempted in this work. 
This was also the aim of the recent work by 
Marigo et al. (\cite{Ma03}), who compute Pop III massive stellar models with
rotation, assuming solid-body rotation. In this very interesting piece of work,
they mainly study the effects of reaching the break-up limit. However,
since they did not include the rotational mixing of the chemical elements, they
could not explore the effects of rotation on the internal chemical composition of the stars.
Also, solid body rotation is just the extreme case of coupling the internal
rotation, which ignores the physics and timescales of the internal transport.
In the present models, the transport of both the angular momentum and the chemical species are treated
in a consistent way, and, as we shall see, rotational mixing has very important consequences on both
the stellar yields and the mass loss rates.

In Sect.~2, we briefly recall the main physical ingredients of the models. 
The evolutions of fast--rotating
massive star models at very low $Z$ are described in Sect. 3. 
The evolution of the internal chemical composition is the subject of Sect.~4,
while the ejected masses of various isotopes are presented in Sect.~5. 
The case of the CEMP stars is discussed in Sect.~6.
Section~7 summarises the main results and raises a few
questions to be explored in future works.

\section{Physical ingredients}

The computation of our models was done with the Geneva evolution code. 
The opacities were taken from Iglesias \& Roger
(\cite{igl96})
and complemented at low temperatures by the molecular opacities of
Alexander (\url{http://web.physics.twsu.edu/alex/wwwdra.htm}). 
The nuclear reaction
rates were based on the NACRE data basis (Angulo \& al. \cite{ang99}). The treatment of
rotation included the hydrostatic effects described in Meynet \& Maeder (\cite{MMI}) and
the effects of rotation on mass loss rates according to Maeder \& Meynet (\cite{MMVI}).
In particular, we accounted for the wind anisotropies induced by rotation as in
Maeder (\cite{MaIV}). Meridional circulation was implemented according to Maeder \& Zahn (\cite{mz98}), but
including the new $D_{\rm h}$ coefficient as described in Maeder (\cite{M03}).
Roughly compared to the old $D_{\rm h}$, the new one tends to reduce the size of
the convective core and to allow larger enrichment of the surface in CNO--processed
elements. The reader is referred to these papers for a detailed description of
the effects. The convective instability was treated according to the
Schwarzschild criterion without overshooting. The radiative mass loss rates are from
Kudritzki \& Puls (\cite{kudpul00}) when $\log T_{\rm eff} > 3.95$ and from
de Jager et al.~(\cite{Ja88}) otherwise. The mass loss rates depend 
on metallicity as $\dot{M} \sim (Z/Z_{\odot})^{0.5}$, where
$Z$ is the mass fraction of heavy elements at the surface
of the star. As we shall see, this quantity may change during the evolution of the star.

%**************
A specific treatment for mass loss was applied at break-up.
At break-up, the mass loss rate adjusts itself in such a way that an
equilibrium is reached between the two following opposite effects. 1) The radial
inflation due to evolution, combined with the growth of the surface velocity due to the
internal coupling by meridional circulation, brings the star to break-up, and thus some
amount of mass at the surface is no longer bound to the star. 2) By removing
the most external layers,
mass loss brings the stellar surface down to a level in the star that
is no longer critical. Thus, at break-up, we should adapt the mass loss rates, in order
to maintain the surface layers at the break-up limit.
In practice, however, since the critical limit contains mathematical
singularities, we considered that during the break-up phase, the mass loss rates should be such
that the model stays near a constant fraction (for example, 0.98) of the limit.
%**************
%A specific treatment for mass loss has been applied at break-up.
%We consider that at break-up the mass loss rate adjusts itself in such a way that an
%equilibrium is reached between the two following opposite effects: 1) the radial
%inflation due to evolution, combined with the growth of the surface velocity due to the
%internal coupling by meridional circulation, brings the star to break-up, and thus some
%amount of mass at the surface is no longer bound to the star. 2) Mass loss, by removing
%the most external layers, brings the stellar surface down to a level in the star which
%is no longer critical. We thus adapt mass loss rates in such a way that the
%super-critical layers are removed, what places the stellar surface at the critical
%break-up limit. In practice, since the critical limit contains mathematical
%singularities, we consider that during the break-up phase, the mass loss rates should be such
%that the model stays at a constant fraction (for example 0.98) of the limit.
At the end of the MS
phase, the stellar radius inflates so rapidly that meridional circulation is unable to
continue to ensure the internal coupling, and the break-up phase ceases naturally.

In this first exploratory work, we focused our attention on stars with initial masses
of 60 M$_\odot$ and 7 M$_\odot$ in order to gain insight into the
properties of both massive and AGB stars at low $Z$.
The evolution was computed until the end of the core helium burning phase (core carbon burning phase
in the case of the 60 M$_\odot$ rotating model at $Z=10^{-8}$).
Two metallicities were considered for the 60 M$_\odot$ models: $Z=10^{-8}$ 
%(corresponding to [Fe/H]$\approx$-6.1
%if we consider for the solar metallicity $Z=0.0122$ given by Asplund et al. \cite{AS05})
and $Z=10^{-5}$. Only this last metallicity was considered for the 7 M$_\odot$ model.
%([Fe/H]$\approx$-3.1). 
%***********
Of course we do not know if stars
with $Z=10^{-8}$ have ever formed; however,
it is not possible at the present time to exclude such a possibility.
Indeed it might be that the first star generations produce very little amounts
of heavy elements, due to the strong fallback of ejected material onto black holes at the end of their lifetimes.
Moreover, as we shall see, the behaviour of the $Z=10^{-5}$ and $10^{-8}$ massive star models
are qualitatively similar, indicating that the evolutionary scenarios explored here might apply to a
broad range of initial metallicities.
%Moreover, we think that it is interesting to explore in addition to the
%Pop III rotating models (which will be discussed in a forthcoming paper), a range
%of metallicities, since a little amount of metals may already have important
%consequences. 
%***********

The initial mixture of heavy elements was taken as equal to the one used
to compute the opacity tables (Iglesias \& Roger
\cite{igl96}, Weiss alpha-enhanced elements mixture).
The initial
composition for models at $Z=10^{-8}$  is given in Table~\ref{tbl-0}.
The models at $Z=10^{-5}$ have the same initial mixture of heavy elements.
More precisely, the mass fractions for all the isotopes heavier than $^{4}$He
were multiplied by $10^3$ (=$10^{-5}/10^{-8}$).

\begin{table}
\caption{Initial abundances in mass fraction for models at
$Z=10^{-8}$.} \label{tbl-0}
\begin{center}\scriptsize
\begin{tabular}{cc}
\hline
             &                              \\
 Element     &    Initial abundance         \\
             &                              \\
\hline
             &                              \\
   H         &        0.75999996            \\
 $^3$He      &        0.00002554            \\
 $^4$He      &        0.23997448            \\
 $^{12}$C    &         7.5500e-10           \\
 $^{13}$C    &         0.1000e-10           \\
 $^{14}$N    &         2.3358e-10           \\
 $^{15}$N    &         0.0092e-10           \\
 $^{16}$O    &         67.100e-10           \\
 $^{17}$O    &         0.0300e-10           \\
 $^{18}$O    &         0.1500e-10           \\
 $^{19}$F    &         0.0020e-10           \\
 $^{20}$Ne   &         7.8368e-10           \\
 $^{21}$Ne   &         0.0200e-10           \\
 $^{22}$Ne   &         0.6306e-10           \\
 $^{23}$Na   &         0.0882e-10           \\
 $^{24}$Mg   &         3.2474e-10           \\
 $^{25}$Mg   &         0.4268e-10           \\
 $^{26}$Mg   &         0.4897e-10           \\
 $^{27}$Al   &         0.1400e-10           \\
 $^{28}$Si   &         3.2800e-10           \\
 $^{56}$Fe   &         3.1675e-10           \\
             &                              \\
 \hline
             &                              \\
\end{tabular}
\end{center}

\end{table}

Nothing is known on the rotational velocities of such stars.
However, there are
some indirect indications that stars at lower $Z$ could have 
higher initial rotational velocities:
1) realistic simulations of the formation of the first stars in the Universe
show that the problem of the dissipation of the angular momentum is
more severe at very low $Z$ than at the solar $Z$. Thus
these stars might begin their evolution 
with a higher amount of angular momentum (Abel et al.~\cite{Ab02}).
2) There are some observational hints that
the distribution of initial rotation might 
contain more fast rotators at lower $Z$ (Maeder et al.~\cite{MG99}).
3) Even if stars begin their life on the ZAMS with the same total amount of
angular momentum 
%(or the same value
%of $\upsilon_{\rm ini}/\upsilon_{\rm crit}$) 
for all metallicities, 
then the stars at lower metallicity
rotate faster as a consequence of their smaller radii.

The three arguments listed above would favour the choice
of a higher value for the initial rotational velocity 
than those adopted for solar models. To choose this value we proceeded 
in the following way. First we supposed that the stars begin their
evolution on the ZAMS with approximately the same angular momentum
content, whatever the metallicity. At solar metallicity, observation provides
values for the mean observed rotational velocity on the MS phase 
(around 200 km s$^{-1}$ for OB stars). Stellar models allowed us to estimate
the initial angular momentum required to achieve such values 
(around 2.2--2.5 10$^{53}$~g~cm$^2$~s$^{-1}$). Adopting
such an initial value of the angular momentum, we found that 
a 60 M$_\odot$ stellar model at $Z = 10^{-8 }$ has
a velocity on the ZAMS of 800 km s$^{-1}$. This is the value
of the initial velocity we adopt in the present work.

\section{Evolution of a massive rotating star at very low metallicity}

\subsection{Rotation and mass loss during the main sequence phase}

\begin{figure}
\resizebox{\hsize}{!}{\includegraphics[angle=-90]{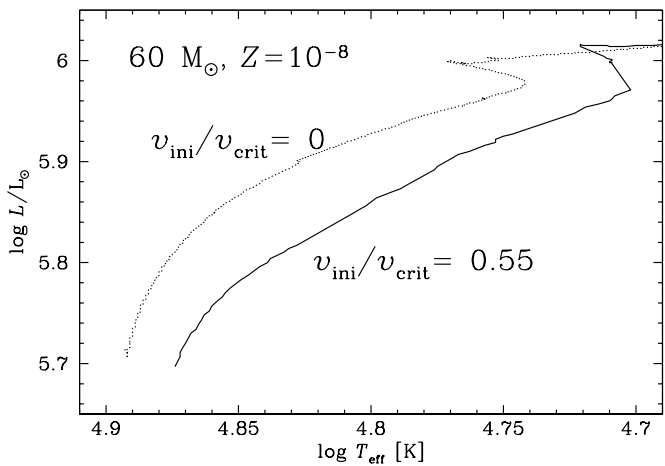}}
%\resizebox{\hsize}{!}{\includegraphics{ooc.jpg}}
\caption{Evolutionary tracks in the HR diagram for 60 M$_\odot$ stellar models
at $Z=10^{-8}$.}
\label{dhrm8}
\end{figure}

Figure~\ref{dhrm8} shows the evolutionary tracks during the main sequence (MS) phase 
for the 60 M$_\odot$ stellar models at $Z=10^{-8}$. Table~\ref{tbl-1} presents some properties of the models
at the end of the core H- and He-burning phases.
%Columns 1 and 2 give the initial mass and the initial metallicity.
%The initial velocity $v_{\rm ini}$ and the 
%mean equatorial rotational velocity $\overline{v}$ during the MS phase
%are indicated in columns 3 and 4 respectively.
%This latter quantity is defined as in Meynet \& Maeder~(\cite{MMV}). 
%The H--burning lifetimes $t_H$,
%the masses $M$, the equatorial velocities $v$, the helium surface abundance $Y_s$ and the 
%surface ratios (in mass fraction) N/C and N/O at the end of the H--burning phase are given in columns 5 to 10.
%Columns 11 to 17 present some characteristics of the stellar models at the end of the He--burning phase,
%$t_{He}$ is the He--burning lifetime. 
From Fig.~\ref{dhrm8}, we see that 
rotation produces a small shift of the tracks
toward lower luminosities and $T_{\rm eff}$. This effect is due to both
atmospheric distortions (note that surface--averaged effective temperatures
are plotted in Fig.~\ref{dhrm8} as explained in Meynet \& Maeder~\cite{MMV})
and to the lowering of the effective gravity 
(see e.g. Kippenhahn and Thomas \cite{KippTh70}; Maeder and Peytremann \cite{MP70}; 
Collins and Sonneborn \cite{co77}). The MS lifetime of the rotating model is enhanced
by 11\%.

These results show that rotation does not
affect the UV outputs of very metal--poor 
massive stars much (the UV outputs of the first
massive star generations might contribute a lot to the reionization of the early Universe,
see {\it e.g.} Madau \cite{Mad03}). 
Only if a significant fraction of primordial stars would rotate
so fast that they follow the path of homogeneous evolution (Maeder~\cite{Ma87}), could rotation
increase the ionizing power. In that case, the star would remain in the blue part of the HR diagram
and would have a much longer lifetime. 

\begin{figure}
\resizebox{\hsize}{!}{\includegraphics[angle=-90]{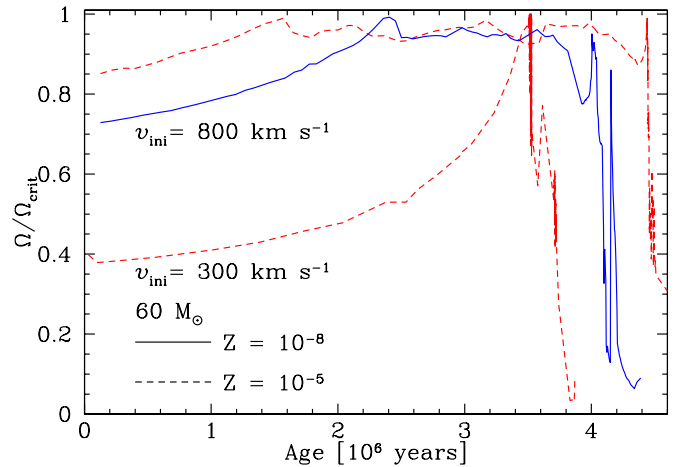}}
%\resizebox{\hsize}{!}{\includegraphics{ooc.jpg}}
\caption{Evolution of $\Omega/\Omega_{\rm crit}$ at the surface of 60~M$_\odot$ models at
$Z=10^{-8}$ (continuous line) and $Z=10^{-5}$ (upper dashed line) with $\upsilon_{\rm
ini}$=~800 km s$^{-1}$. The case of the 60 M$_\odot$ model
at $Z=10^{-5}$ with $\upsilon_{\rm ini}$=~300 km s$^{-1}$
from Meynet \& Maeder~(\cite{MMVIII}) is also shown (lower dashed line).}
\label{ooc}
\end{figure}

Figure~\ref{ooc} shows the evolution of the ratio $\Omega/\Omega_{\rm crit}$ 
at the surface during the MS phase. At $Z=10^{-8}$, the model with $\upsilon_{\rm ini}=800$ km~s$^{-1}$
reaches the break-up limit when the mass fraction of hydrogen at the centre $X_{\rm c} \simeq$~0.40.
The star stays at
break-up for the remaining of its MS life with an enhanced mass loss rate.
As a consequence, the model ends its MS life with 57.6~M$_\odot$, having lost 4\% of
its initial mass. 
Despite the star staying in the vicinity of the break-up limit during an important part of
its MS lifetime, it does not lose very large amounts of mass, 
due to the fact that only the outermost layers of 
the stars are above the break--up limit and are ejected. These layers have low density and thus contain little mass.

A model with the same initial velocity, but with a metallicity three orders
of magnitude higher, reaches the break--up limit very early in the MS phase, when $X_{\rm c} \simeq$~0.56.
This comes from the fact that when the metallicity increases, a given value of the initial 
velocity corresponds to a higher initial value
of the $\upsilon_{\rm ini}/\upsilon_{\rm crit}$ ratio.
The model at $Z=10^{-5}$
ends its MS life with 53.8~M$_\odot$, having lost 10\% of
its initial mass. 

\begin{table*}
\caption{Properties of the stellar models at the end of the H-
and He-burning phases. 
$M_{\rm ini}$ is the initial mass, $Z$ the initial metallicity,
$v_{\rm ini}$ the initial velocity, $\overline{v}$ the 
mean equatorial rotational velocity during the MS phase defined as in Meynet \& Maeder~(\cite{MMV}),
$t_{\rm H}$ the H--burning lifetimes, $M$ the actual mass of the star, $v$ the actual rotational
velocity at the stage considered, $Y_{\rm s}$
the helium surface abundance in mass fraction. N/C and N/O are the ratios of nitrogen to carbon,
respectively, of nitrogen to oxygen at the surface of stars in mass fraction; C, N, O are the abundances
of carbon, nitrogen, and oxygen at the surface in mass fractions.
The numbers in parentheses indicate the power of ten, {\it i.e.}, 7.54(-10)=7.54 $\times 10^{-10}$.} \label{tbl-1}
\begin{center}\scriptsize
\begin{tabular}{cccc|cccccc|ccccccc}
\hline
    &         &         &         &         &     &      &      &      &     &      &      &      &         &   & &         \\

\multicolumn{4}{c|}{ } & \multicolumn{6}{|c|}{End of H--burning} &\multicolumn{7}{|c}{End of He--burning}      \\
    &         &         &         &         &     &      &      &      &     &      &      &      &         &   & &         \\
$M_{\rm ini}$   &  $Z$ & $v_{\rm ini}$ & $\overline{v}$ & $t_{\rm H}$ & $M$ & $v$ & $Y_{\rm s}$ & N/C & N/O &  $t_{\rm He}$ & $M$ & $v$ & $Y_{\rm s}$ & C &  N & O \\
M$_\odot$    &     & ${\rm km} \over {\rm s}$ & ${\rm km} \over {\rm s}$ & Myr & M$_\odot$ & ${\rm km} \over {\rm s}$ &  &  &  &  
Myr  & M$_\odot$ & ${\rm km} \over {\rm s}$ & &  &   &  \\
    &         &         &         &         &     &      &      &      &     &      &      &      &         &    & &        \\
\hline
\hline
\multicolumn{4}{c|}{ } &  \multicolumn{6}{|c|}{ } &\multicolumn{7}{|c}{ }       \\
60  &  10$^{-8}$ &   0 &   0 & 3.605  &  59.817 &   0 & 0.24 & 0.31 & 0.03 &  0.292 & 59.726 & 0 & 0.24 & 7.54(-10) & 2.34(-10) & 6.71(-9)     \\
60  &  10$^{-8}$ & 800 & 719 & 4.004  &  57.624 &  591 & 0.27 & 103 & 9.77 &  0.522 & 23.988 & 0.02  & 0.76 & 1.97(-4) & 1.02(-2) & 2.85(-4)      \\
%60  &  10$^{-5}$ &   0 & 0   & 3.883 & 59.787  &  0   & 0.23 & 0.31 & 0.03 & 0.332 & 59.699 & 0  & 0.23 & %1.00 &  1.00 & 59.57 & 0   & 0.23 & 1.00 & 1.00   \\
60  &  10$^{-5}$ & 800 & 636 & 4.441  &  53.846 &  567 & 0.34 &  40 & 0.82 &  0.544 & 37.280 & 0.57  & 0.80 & 5.06(-5) & 2.07(-3) & 4.24(-5)     \\
% 7  &  10$^{-5}$ & 800 & 377 & 44.403 &   6.999 &  400 & 0.24 & 167 & 0.50 &  6.237 &  6.988 & 1.70  & 0.33 & 2.73(-3) & 3.00(-3) & 2.95(-3)     \\
    &         &         &         &         &     &      &      &      &     &      &      &      &         &    & &        \\
\hline
\multicolumn{17}{ }{ }       \\
\end{tabular}
\end{center}

\end{table*}

In
order to discuss the effects of a change of rotation, it is
interesting to compare this last result
with the one obtained in Meynet \& Maeder~(\cite{MMVIII}) for a 60~M$_\odot$ model at $Z=~10^{-5}$ with
$\upsilon_{\rm ini}$=~300 km s$^{-1}$.  
This last model reaches the break-up velocity much later, only when $X_{\rm c}
\simeq$~0.01. At $Z=10^{-5}$, the velocity 300 km s$^{-1}$ thus appears as the lower limit
for the initial rotation,
allowing a 60 M$_\odot$ star to reach the break-up limit during its MS phase.
The 60 M$_\odot$ star ends its MS life with 59.7~M$_\odot$, having lost only 0.5\% of its
initial mass.
Note that the 2002 models were computed with
slightly different physical ingredients than those used to compute the models
discussed in this paper (different expression
for $D_{\rm h}$ and for the mass loss rates).
However, at these very low metallicities, radiatively driven winds remain quite modest, and 
the transport of the angular
momentum, mainly driven by the meridional circulation, does not depend much
on the expression of $D_{\rm h}$.

\begin{figure}
\resizebox{\hsize}{!}{\includegraphics[angle=-90]{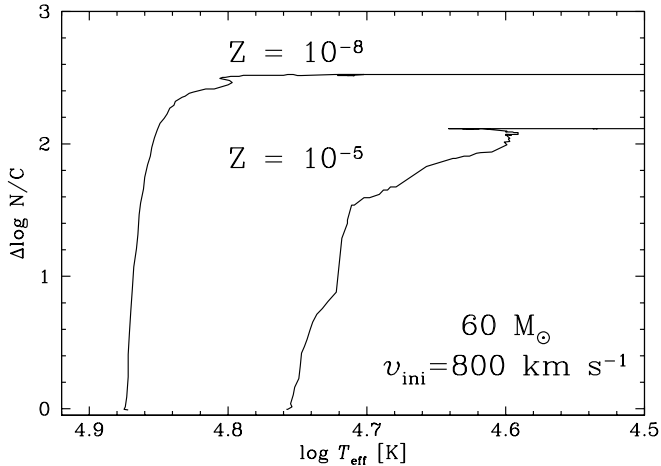}}
%\resizebox{\hsize}{!}{\includegraphics{rgb.jpg}}
\caption{Evolution as a function of $\log T_{\rm eff}$ of the excess at the surface for the
ratio N/C (expressed in dex) compared to the initial ratio.
N and C are the abundances of nitrogen and carbon
at the surface of the star.}
\label{nc}
\end{figure}

During the MS phase, the surface of the rotating stars is 
enriched in nitrogen and depleted in carbon as a result of rotational mixing.
Figure~\ref{nc} shows that
the N/C ratios are enhanced by more than two orders of magnitude 
at the end of the H-burning phase.
More precisely, at the surface of the $Z=10^{-8}$ model, 
nitrogen is enhanced by a factor 27 and carbon decreased
by a factor 12. All other physical ingredients 
being the same, one also sees that the model with 
the lowest metallicity is also the one with the greatest surface enrichments. This well agrees
with the trend already found in our previous works (Maeder \& Meynet~\cite{MMVII};
Meynet \& Maeder \cite{MMVIII}), which
results from 
the steep gradients of angular velocity that build up 
in very metal--poor stars that favours shear mixing. 
Let us emphasise, however, that during the MS phase,
the arrival of CNO--processed material
at the surface
does not change the 
metallicity of the outer layers.
Indeed, rotational mixing brings nitrogen to the surface 
but depletes carbon and oxygen, keeping the sum of CNO elements constant,
and the metallicity as well. Thus during the MS phase, 
the enhanced mass loss rates
undergone by the rotating models are entirely due 
to the mechanical effect of the centrifugal force.
As we shall see, this is no longer the case 
during the core He-burning phase.

\begin{figure}
\resizebox{\hsize}{!}{\includegraphics[angle=-90]{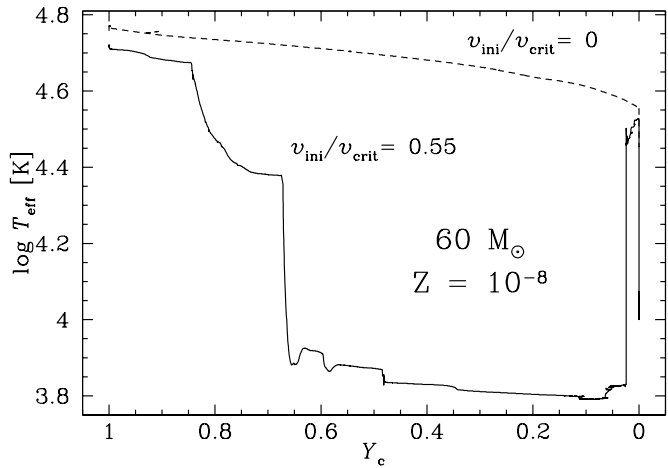}}
%\resizebox{\hsize}{!}{\includegraphics{rgb.jpg}}
\caption{Evolution of $\log T_{\rm eff}$ as a function of $Y_{\rm c}$, the mass fraction of
$^4$He at the centre, for a non-rotating (dashed line) and rotating (continuous line)
60~M$_\odot$ model at $Z=~10^{-8}$.}
\label{rgb}
\end{figure}

\subsection{Rotation and mass loss during the post main-sequence phases}

%Starting with an initial ratio of 0.87, the $Z=~10^{-5}$ model reaches the
%critical velocity very early in the MS, when $X_{\rm c} \simeq$~0.63. 

%The natural
%acceleration counterbalanced by enhanced mass loss (as described above) maintains it at
%break-up limit for the remaining two third of its MS life. The needed adaptation of the
%mass loss to the break-up situation leads to an enhancement of 1-2 orders of magnitude,
%from $\log \dot{M} \ssim -7$ at the beginning of the MS to $-6 \ssim \log \dot{M} \ssim
%-5$ near break-up. The model ends its MS life with 54.2~$M_\odot$, having lost 10\% of
%its initial mass. Comparatively, the model at $Z=~10^{-8}$ begins the MS with an expected lower
%$\Omega/\Omega_{\rm crit}$ ratio of 0.77. It reaches
%the break-up limit later in the MS, when $X_{\rm c} \simeq$~0.40 but stays also at
%break-up limit until the end of the MS, losing about 4\% of its initial mass. It is
%interesting to compare our results to the 60~$M_\odot$ model at $Z=~10^{-5}$ and
%$\upsilon_{\rm ini}$=~300 km/s computed in Paper VIII (hereafter the slow model), in
%order to underscore the effects of fast rotation. We can see that starting with an
%initial ratio of 0.38, this slow model reaches the break-up velocity only at $X_{\rm c}
%\simeq$~0.01 and ends its MS life with 59.7~$M_\odot$, having lost only 0.5\% of its
%initial mass. Thus, these various models offer a consistent picture.

\begin{figure*}
\resizebox{\hsize}{!}{\includegraphics[angle=-90]{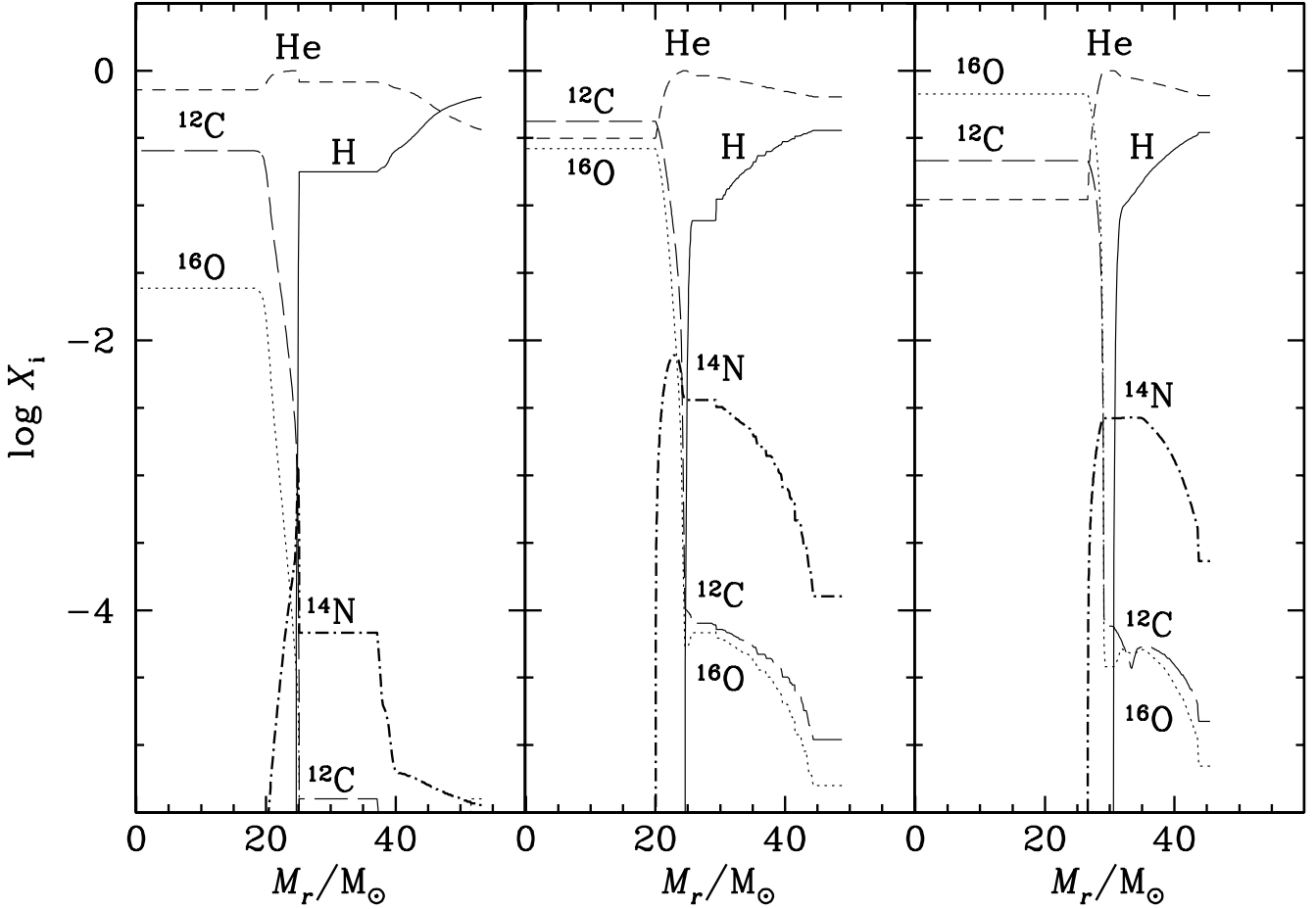}}
%\resizebox{\hsize}{!}{\includegraphics{rgb.jpg}}
\caption{Chemical composition of a 60 M$_\odot$ stellar model at $Z=10^{-5}$ with
$\upsilon_{\rm ini}$= 800 km s$^{-1}$ when it evolves from the blue to the red
part of the HR diagram. The model shown in the left panel has
$\log L/{\rm L}_\odot = 6.129$ and $\log T_{\rm eff}=4.243$; in the middle
panel, it has $\log L/{\rm L}_\odot = 6.130$ and $\log T_{\rm eff}=4.047$; in the
right panel, it has $\log L/{\rm L}_\odot = 6.145$ and $\log T_{\rm eff}=3.853$.}
\label{travers}
\end{figure*}

\begin{figure*}
\resizebox{\hsize}{!}{\includegraphics[angle=-90]{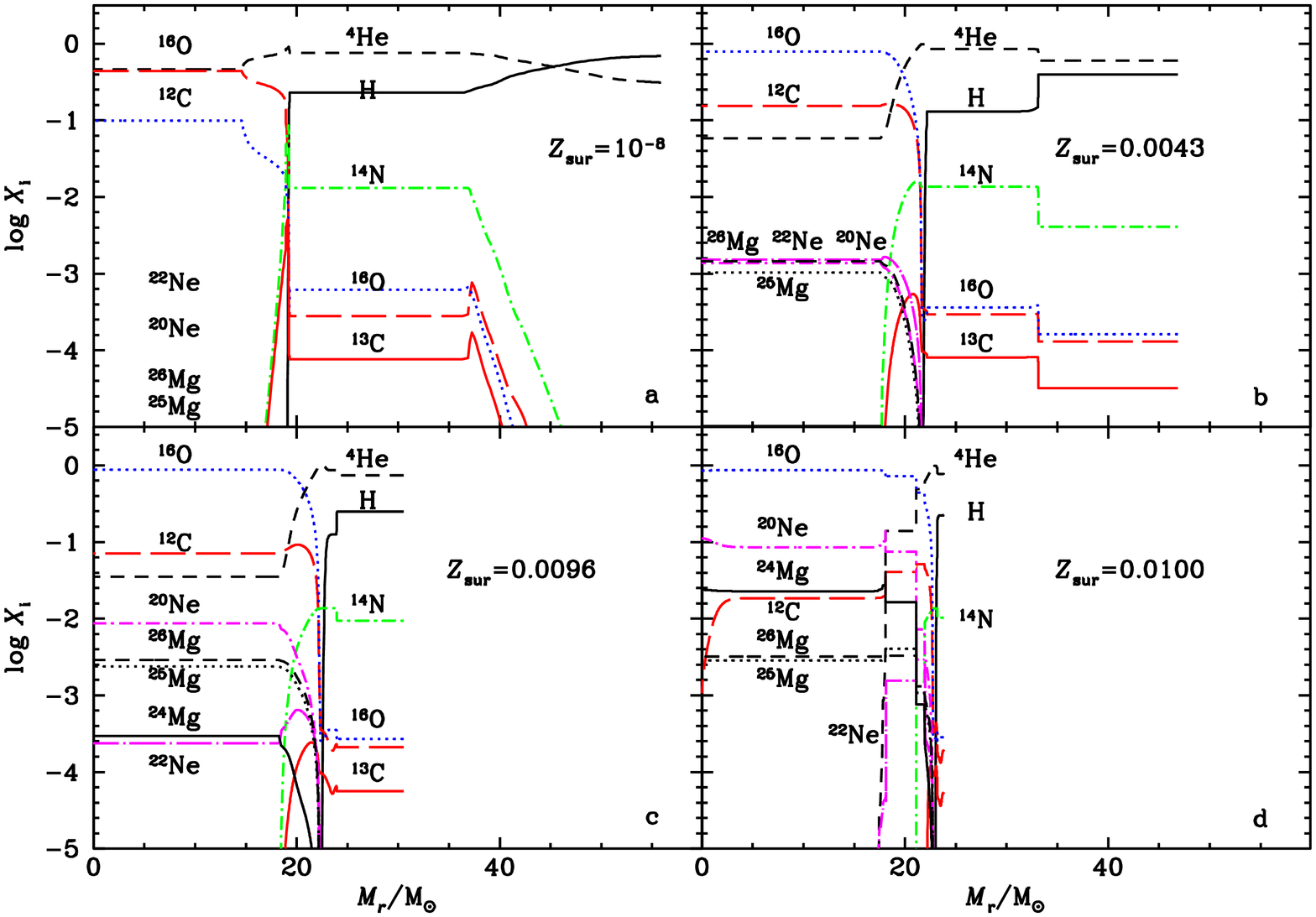}}
\caption{Variations of the abundances (in mass fraction) as a function of the Lagrangian mass
within a 60~M$_\odot$ star with
$\upsilon_{\rm ini}$=~800~km~s$^{-1}$ and $Z=10^{-8}$. The four panels show the chemical
composition at four different stages at the end of the core He-burning phase:
in panel {\bf a)} the model has a mass fraction of helium at the centre, $Y_{\rm c}$=~0.11
and an actual mass $M$=~54.8~M$_\odot$ - {\bf b)} $Y_{\rm c}$=~0.06,
$M$=~48.3~M$_\odot$ - {\bf c)} $Y_{\rm c}$=~0.04, $M$=~31.5~M$_\odot$
- {\bf d)} End of the core C-burning phase, $M$=~23.8~M$_\odot$. The actual surface metallicity $Z_{\rm surf}$
is indicated in each panel.}
\label{abond}
\end{figure*}

Rotational mixing that occurs during the MS phase and 
still continues to be active during the core He-burning 
phase deeply modifies the internal chemical composition 
of the star. This has important consequences during the 
post-MS phases and deeply changes the evolution 
of the rotating models with respect to their non--rotating 
counterparts. Among the most striking differences, one notes the following:

\begin{itemize}
\item 1) Rotation favours redwards evolution in the 
HR diagram as was already shown by
Maeder \& Meynet (\cite{MMVII}), and as illustrated 
in Fig.~\ref{rgb}. One sees that 
the non-rotating model remains on the blue side during 
the whole core He-burning phase, while
the 800 km s$^{-1}$ model at $Z$ = 10$^{-8}$
starts its journey toward the red side of the HR diagram early in the core helium burning
stage, when $Y_{\rm c} \simeq 0.67$
($Y_{\rm c}$ is the mass fraction of helium
at the centre of the star model). The same is true for the corresponding model at 
$Z$ = 10$^{-5}$. Let us recall that this behaviour is linked to 
the rapid disappearance of the intermediate convective 
zone associated to the H-burning shell (see Fig.~\ref{travers} and 
Maeder \& Meynet~\cite{MMVII}). 
%In rotating models,
%convection is less favoured in this zone because the H-shell is less active. 
%The lower activity of the H-shell comes from the facts that in rotating models
%the helium cores are more massive and thus contribute more to the total luminosity of the star
%and from helium diffusion which
%decreases  the quantity of hydrogen in that zone.
\item 2) Redwards evolution enhances the mass loss. 
In the cases of our 60 M$_\odot$ stellar models,
it brings the stars near the Humphreys--Davidson limit, 
{\it i.e.}, near $\log L/{\rm L}_\odot=6$ and $T_{\rm eff}$ in a broad range around $10^4$ K.
Near this limit, the mass loss rates (here from de Jager et al.~\cite{Ja88}) become very important. 
%At solar metallicity one has in this region
%mass loss rates of the order of $\log(-\dot M)=-3)$
%where $\dot M$ is in M$_\odot$ per year.
For instance,
the model represented 
in the left panel of Fig.~\ref{travers} ($\log L/{\rm L}_\odot= 6.129,\  \log T_{\rm eff}= 4.243$) is still far
to the left hand side of the Humphreys-Davidson limit. Its mass loss rate
is 
$\log (-\dot M)=-5.467$, where $\dot M$ is expressed in M$_\odot$ per year.
The model in the right panel ($\log L/{\rm L}_\odot= 6.145,\  \log T_{\rm eff}= 3.853$) is in the vicinity of the
Humphreys-Davidson limit. Its mass loss rate is equal to -4.616, {\it i.e.}, more than seven times higher.
During this
transition, the overall metallicity at the surface 
does not change and remains equal to the initial 
one (here $Z_{\rm ini}=0.00001$). We observe a similar 
transition in the case of the $Z=10^{-8}$ stellar model. 
\item 3) During the core He-burning phase, primary nitrogen is synthesized in the H-burning shell, due
to the rotational diffusion of carbon and oxygen produced in the helium core into the H-burning shell 
(Meynet \& Maeder \cite{MMVIII}). This is illustrated well in 
Fig.~\ref{travers} for the $Z=10^{-5}$ rotating model and in Fig.~\ref{abond}
for the model at $Z=10^{-8}$. 
\item 4) In contrast to what happens during the MS phase,
rotational mixing during the core He-burning phase
induces large changes in the surface metallicity.
These changes occur only at the end of the core He-burning phase,
although  the conditions for their apparition result from the mixing
that occurs during the whole core He-burning phase. Indeed,
rotational mixing progressively enriches the outer radiative zone
in CNO elements, thus enhancing its opacity slowly. 
When, in the $Z$ = 10$^{-8}$ stellar model, the abundance of nitrogen in the outer layers becomes approximately
$10^{-8}$ in mass fraction ({\it i.e.}, has increased
by two orders of magnitude with respect to the initial value),
these outer layers become convective. The outer convective zone
then rapidly deepens in mass and
dredges up newly synthesized elements to the surface.
From this stage onwards,
the surface metallicity increases in a spectacular way,
as can be seen in Fig.~\ref{abond}. For instance, the rotating
60 M$_\odot$ at $Z=10^{-8}$ has a surface metallicity of $10^{-2}$ at the end of its lifetime,
{\it i.e.}, similar
to that of the Large Magellanic Cloud !
\item 5) The consequence of such large surface enrichments 
on the mass loss rates remains to be studied in detail 
using models of stellar winds with the appropriate physical 
characteristics (position in the HR diagram and chemical composition). 
In the absence of such sophisticated models, we applied 
the usual rule here, namely $\dot M(Z)=(Z/Z_\odot)^{1/2}\dot M(Z_\odot)$, where $Z$
is the metallicity of the outer layers.  
With this prescription, the surface enhancement of the metallicity 
is responsible for the large decrease in the stellar mass 
that can be seen in Fig.~\ref{abond}.
%With a dependence of the $\dot M$-rates like $Z^{0.8}$, the effects
%shown here would be larger.
\item 6) During the late stages of the core helium--burning phase, 
as a result of mass loss and mixing, the star 
may evolve along a blue loop in the HR diagram (see Fig.~\ref{rgb}). 
When the star evolves bluewards,
the global stellar contraction brings the outer convective zone,
which evolves like a solid body rotating shell to break-up
(Heger \& Langer \cite{He98}). 
At this stage of the evolution, the luminosity is not far from
the Eddington limit and the star may reach the $\Omega\Gamma$-limit
(Maeder \& Meynet \cite{MMVI}). This multiplies
the mass loss rates by very large factors. 
\item 7) During the last 24 000 years of its lifetime, the model
presents abundance patterns characteristic of WNL stars at its surface.
\item 8) As a result of mixing and mass loss rates, the duration
of the core He-burning phase
is much longer in the rotating model. The present
60 M$_\odot$ model with $\upsilon_{\rm ini}=800$ km s$^{-1}$ at $Z=10^{-8}$ has
a helium--burning lifetime that is $\sim$80\% longer than the corresponding lifetime of the
non-rotating model (see Sect.~4 below for more explanations).
\end{itemize}  

The different effects described above are all due to the mixing induced by rotation and
they all tend to enhance the quantity of mass lost by stellar winds. 
The rotating 60~M$_\odot$ at $Z=10^{-8}$ loses about 36 M$_\odot$ during 
its lifetime. About 2 M$_\odot$ are lost due to break-up during the MS phase,
$\sim$ 3 M$_\odot$ are lost when the star is 
in the red part of the HR diagram (with surface metallicity equal to the initial one), 27 M$_\odot$ are lost
due to the effect of the enhancement of the surface metallicity, the remaining 4 M$_\odot$ are lost
when the star evolves along the blue loop and reaches the $\Omega\Gamma$-limit.
One sees that, by far, the most important effect is due to the increase in the surface metallicity.

Paradoxically the corresponding model at higher 
metallicity ($Z=10^{-5}$) loses less mass (a little less than 40\% of the total mass). 
This can be understood from the following facts: first less 
primary nitrogen is synthesized due to slightly less 
efficient chemical mixing when the metallicity increases, thus the surface
does not become as metal rich as in the model at $Z=10^{-8}$.
Second and for the same reason as above, the outer convective zone does not deepen as far 
as in the more metal--poor model. 
These two factors imply that the maximum 
surface metallicity reached in this model, 
which is equal to 0.0025, is about a factor 4 
below the one reached by the $Z=10^{-8}$ model. 
Finally, the blue loop does not extend that far into the 
blue side, and the surface velocity always remains well below 
the break-up limit during the whole blueward excursion.

In order to investigate to what extent the behaviour described above depends
on the physical ingredients of the model, we
compare the present results with those of a rotating model ($\upsilon_{\rm ini}$= 800 km s$^{-1}$) of 
a 60 M$_\odot$ star at $Z=10^{-5}$
with a different prescription for the mass loss rates
(Vink et al.~\cite{vink00}, \cite{vink01} instead of Kudritzki \& Puls~\cite{kudpul00}), 
with the Ledoux criterion instead of the Schwarzschild one for determining the size of the convective core,
with a core overshoot of $\alpha=0.2~H_p$
and the old prescription for the horizontal diffusion coefficient $D_{\rm h}$.
This model is described in Meynet et al.~(\cite{Meynetal05}).
In this case, the outer convective zone
deepens farther into the stellar interior and thus produces a greater enhancement of the surface 
metallicity (the same order as the one we obtained in the present $Z=10^{-8}$ 60 M$_\odot$ model).
Higher enhancements of the surface metallicity then induces greater mass loss by stellar winds.
More important than these differences, however,
we shall retain here 
that the results are qualitatively similar to those obtained in our previous models.
In particular,
the mechanism of surface metallicity enhancement 
occurs in both models and appears to be a robust process.

\subsection{Do very metal--poor, very massive stars end their lives as pair--instability supernovae~?}

Might the important mass loss undergone by the rotating models prevent 
the most massive stars from going through pair instability~?
According to Heger \& Woosley (\cite{HW02}), progenitors of pair--instability supernovae
have helium core masses
between $\sim$64 and 133 M$_\odot$. This corresponds to initial masses between about 140 and 260 M$_\odot$.
Thus the question is whether
stars with initial masses above 140 M$_\odot$ can lose a sufficient amount of mass
to have a helium core that is less than about 64 M$_\odot$
at the end of the core He-burning phase. 
From the values quoted above, it would imply the loss of more than
(140-64)=76 M$_\odot$, which represents about 54\%
of the initial stellar mass. 
From the computation performed here, 
one can expect that such a scenario is possible,
where
a 60 M$_\odot$ loses more than 60\% of its initial mass. However, 
more extensive computations are needed to
check whether the rotational mass loss could indeed prevent the most massive stars
from going through this pair instability. Were this the case, it would explain why
the nucleosynthetic signature of pair--instability supernovae is not observed
in the abundance pattern of the most metal--poor halo stars observed up to now.
At least this mechanism could restrain the mass range for the progenitors
of pair--instability supernovae, pushing the minimum initial mass needed for such a scenario to occur to higher values.
Let us also note that the luminosity of the star comes
nearer to the Eddington limit
when the initial mass increases. When rotating, such stars will then encounter
the $\Omega\Gamma$-limit (Maeder \& Meynet~\cite{MMVI}) and very likely undergo strong mass losses.

\section{Evolution of the interior chemical composition}

\begin{figure}
\resizebox{\hsize}{!}{\includegraphics{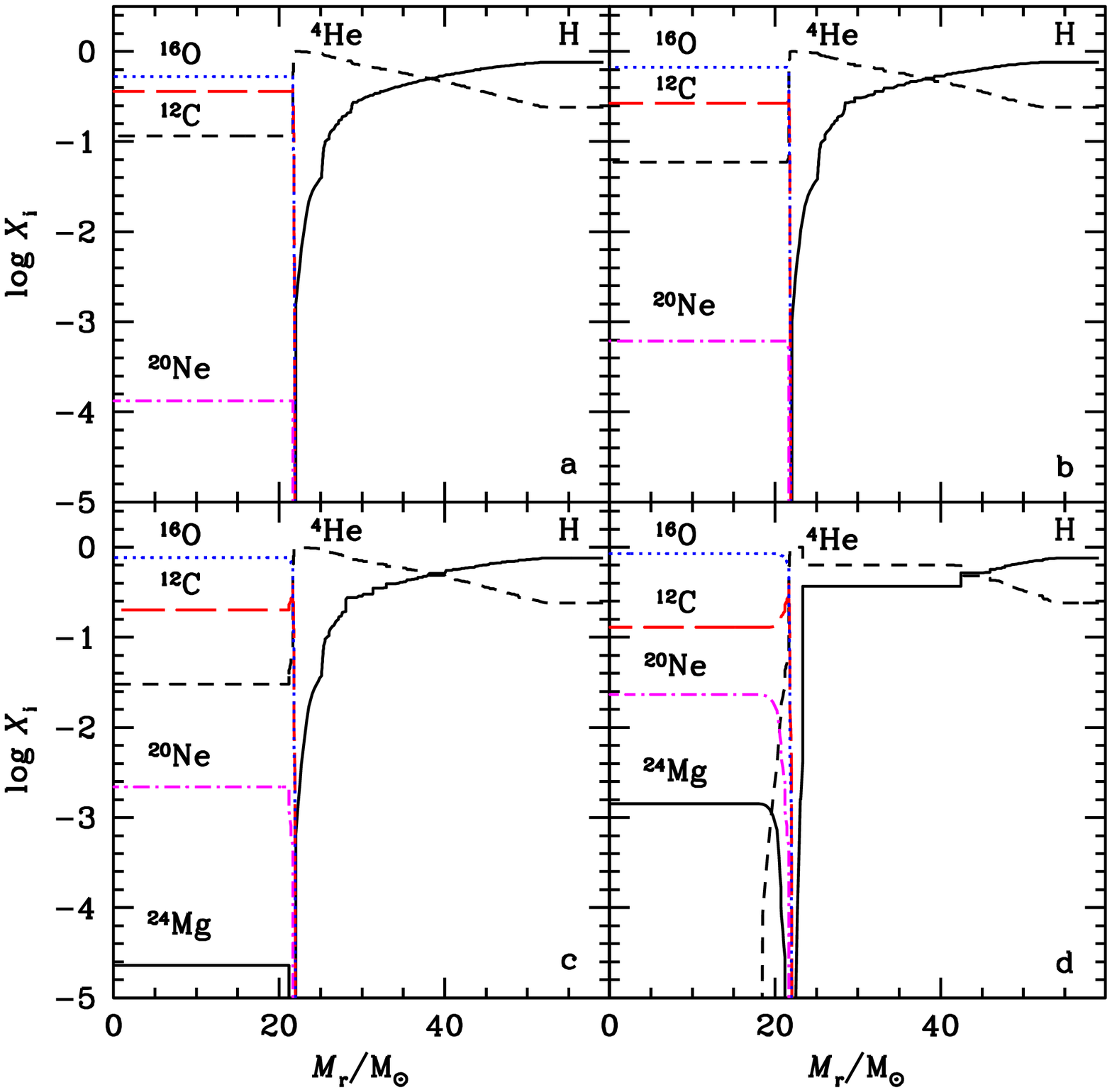}}
\caption{Same as Fig.~\ref{abond} for a 60~M$_\odot$ star with
$\upsilon_{\rm ini}$=~0~km s$^{-1}$ and $Z=10^{-8}$. The four panels shows the chemical
composition at four different stages at the end of the core He-burning phase:
Panel {\bf a)} $Y_{\rm c}$=~0.12, $M$=~59.74~M$_\odot$ - {\bf b)} $Y_{\rm c}$=~0.06,
$M$=~59.74~M$_\odot$ - {\bf c)} $Y_{\rm c}$=~0.03, $M$=~59.73~M$_\odot$
- {\bf d)} $Y_{\rm c}$=~0.0, $M$=~59.73~M$_\odot$. The surface
metallicity is equal to $10^{-8}$ at the four evolutionary stages.}
\label{abond2}
\end{figure}

As discussed above, rotational mixing changes the chemical
composition of stellar interiors in an important way. 
This is illustrated well by Figs.~\ref{abond} and \ref{abond2}, which show
the internal chemical composition of our rotating and 
non-rotating 60 M$_\odot$ stellar models at four different stages
at the end of the core He--burning phase (models at $Z=10^{-8}$).

Comparing panels {\bf a} of Figs.~\ref{abond} and \ref{abond2}, one sees that a large
convective shell is associated to the H-burning shell in the rotating
model, while such a shell is absent in the non-rotating model. 
This contrasts with what happens at the 
beginning of the core He-burning phase, where the
intermediate convective zone associated to the H-burning shell was absent in the rotating model (or at least much smaller),
while in the non-rotating model, the intermediate convective zone was well--developed (see above).
Why is there this difference between the beginning and the end of the core He-burning phase~?
At the beginning of the core He-burning phase, the disappearance of the intermediate
convective shell was a 
consequence of the rotational mixing that operated during 
the core H-burning phase and that brought some freshly synthesized helium
into this region. 
More helium in this region means less hydrogen and also some decrease
in the opacity, both of which inhibit the development of convection
(cf. Maeder \& Meynet~\cite{MMVII}). 
Now, at the end of the core He-burning phase, we have 
He-burning products that are brought into the H-burning shell.
These products, mainly carbon and oxygen, act as catalysts 
for the CNO cycle and make the H-burning shell more active, thus
favouring convection. This mechanism enriches 
this zone not only in primary $^{14}$N, but also in primary $^{13}$C. 
Looking at the H-rich envelope in the non-rotating model at the same stage, 
one sees that all these elements have much lower abundances. Actually they fall
well below the minimum ordinate of the figure.

If one now compares the chemical composition of the 
CO cores when $Y_c\sim 0.11$ (see panels {\bf a} of Figs.~\ref{abond} and \ref{abond2}), 
one notes the following points.
First, the abundances in $^{12}$C, $^{16}$O and $^{20}$Ne are approximately equal
in both the rotating and non-rotating models. This comes from the fact
that the CO core masses are approximately the same in both models. 
On the other hand in the rotating
model, the abundance of $^{22}$Ne is greatly enhanced, 
as the abundances of $^{25}$Mg and $^{26}$Mg. 
The abundance of
$^{22}$Ne results from the conversion of primary $^{14}$N, 
which has diffused
into the He-burning core. Thus the resulting high abundance of $^{22}$Ne
is also of primary origin. 
The isotopes of magnesium are produced by the reactions $^{22}$Ne($\alpha$, $\gamma$)$^{26}$Mg
and $^{22}$Ne($\alpha$, n)$^{25}$Mg.
Their high abundances also result from primary nitrogen diffusion into the He-core.

The CO--core mass (cf. Table~\ref{tbl-2}) in the rotating model is slightly smaller than in the non-rotating one. 
This contrasts to what
happens at higher metallicity, where rotation 
tends to increase the CO--core mass (see Hirschi et al. \cite{Hi04}). Again, this
results from the mechanism of primary nitrogen production, 
which induces a large convective zone associated to the H-burning shell, which then 
prevents this shell from migrating outwards and, thus, the CO core from growing in mass.
Let us recall that in rotating models at solar metallicity, there is no primary nitrogen
production due to the less efficient mixing at higher metallicity (see
Meynet \& Maeder~\cite{MMVIII}); thus there is no increase in the H-burning shell
activity.

In panel {\bf b} of Fig.~\ref{abond}, as explained in the previous section,
one sees the outer convective zone extending inwards 
and bringing CNO elements to the
surface. In panels {\bf c} and {\bf d}, mass--loss efficiently removes these outer layers.
At the corresponding stages
in the non-rotating model, the outer envelope is not enriched in heavy elements and
keeps its mass.

At the end of the He-burning phase (see panels {\bf d}), the abundance of $^{12}$C 
is significantly smaller in the rotating model 
than in the non-rotating one. At the same time, the abundances of $^{20}$Ne and $^{24}$Mg
are significantly greater. This is a consequence of helium diffusion into the He-core
at the end of the He-burning phase. Let us recall that
$^{12}$C is destroyed by alpha capture (to produce $^{16}$O), while $^{20}$Ne and $^{24}$Mg are produced by
alpha captures on, respectively, $^{16}$O and $^{20}$Ne. For what concerns the other isotopes of neon 
and magnesium, one sees that in the rotating models, much higher
abundances of $^{25}$Mg and $^{26}$Mg are reached due to the transformation of the $^{22}$Ne at the end of the 
core helium--burning phase. The neutrons liberated by the $^{22}$Ne($\alpha$,n)$^{25}$Mg reaction can be
captured by iron peak elements, producing some amount of s-process elements (see e.g Baraffe et al.~\cite{Ba92}).
In view of the important changes to the interior chemical composition due to rotation, there is good chance
that the s-process in the present rotating massive star models is quite
different from the one obtained in non-rotating models. This will be examined in later papers.

\section{Chemical composition of the winds and of the supernova ejecta}

%wind
\subsection{Wind composition}

Let us first discuss the chemical composition of the winds. The total mass lost, as well as the quantities
of various chemical elements ejected by stellar winds, are given in Tables~\ref{tbl-2} \& \ref{tbl-3}.
The models at $Z=10^{-5}$ were computed with an extended nuclear reaction network including
the Ne-Na and Mg-Al chains, which is why in Table~\ref{tbl-3} the wind--ejected masses of these elements can
be indicated. The stellar yields - {\it i.e.}, the mass of
an isotope newly synthesized and ejected by the star - can be obtained by subtracting 
the mass of the isotope
initially present in that part of the star\footnote{This quantity may be obtained
by multiplying the initial abundance
of the isotope considered (given in Table~\ref{tbl-0}) by
$m_{\rm ej}$. }
from the ejected masses
given in Tables~~\ref{tbl-2} \& \ref{tbl-3}. 
%For all the heavy elements ejected by
%the rotating models, the ejected masses are nearly identical to the stellar yields.

According to Table~\ref{tbl-2}, the non-rotating model ejects 
only half a percent of its total mass through stellar winds, which is completely negligible. Moreover,
this material has exactly the same chemical composition as does the protostellar cloud from which
the star formed. If, 
at the end of its lifetime, all the stellar material is swallowed by the
black-hole resulting from the star collapse, the nucleosynthetic contribution of such stars would be zero.
In contrast, the rotating models lose more than 60\% of their initial mass through stellar winds.
This material is strongly enriched in CNO elements. Even if all the final stellar mass is
engulfed into a black-hole at the end of
the evolution, the nucleosynthetic contribution of such stars 
remains quite significant.
As already noted above, the corresponding model at $Z=10^{-5}$ loses less mass by stellar winds (see Table~\ref{tbl-3}).
However, the amounts of mass lost remain large and they present strong
enrichments in CNO elements, as in the case of the $Z=10^{-8}$ rotating model. Also $^{23}$Na and $^{27}$Al are somewhat
enhanced in the wind material. 

Other striking differences between the rotating
and non-rotating models concern the $^{12}$C/$^{13}$C, N/C, and N/O ratios (see Tables~\ref{tbl-2} \& \ref{tbl-3}).
The wind of the non-rotating model shows solar ratios ($^{12}$C/$^{13}$C=73, N/C=0.31, and N/O=0.03 in mass fractions). 
The wind of rotating models is characterised by very low $^{12}$C/$^{13}$C ratios, around 4 - 5 
(close to the equilibrium value of the CN cycle) and by very high
N/C (between about 3 and 40) and N/O ratios (between 1 and 36). Thus wind material presents the signature of heavily CNO--processed material.

\subsection{Total ejecta composition (wind and supernova ejecta)}
%wind+sn
%method
In order to estimate the quantity of mass lost at the time of the supernova explosion (if any), it is
necessary to know the mass of the remnant. 
This quantity is estimated with the relation of Arnett (\cite{Ar91})
between the mass of the remnant and the mass of the carbon-oxygen core.
%The masses of the CO cores 
%(m$_{\rm CO}$) of the present stellar models and of the remnant ($m_{\rm rem}$) estimated with the relation
%given by Arnett (\ref{Ar91}) are given in Tables~\ref{tbl-2} and \ref{tbl-3}.
The masses of the different elements ejected are then simply obtained by integrating their
abundance in the final model between $m_{\rm rem}$ (see Tables~\ref{tbl-2} \& \ref{tbl-3}) and the surface. 
Since the evolution of the present models was stopped before the presupernova stage was reached, 
the masses of $^{12}$C and $^{16}$O obtained here
might still be somewhat modified by the more advanced nuclear phases. 
%However, they will not much vary
%as one can convince oneself by comparing 
%the masses of $^{16}$O
%obtained by integration in our models and the masses of this same element obtained
%from the presupernova models of Arnett~(\cite{Ar91}).
%The comparison has to be made
%between models with similar masses of the He-core. We obtain that the quantities given %in Tables~\ref{tbl-2}
%and \ref{tbl-3} do not differ by more than 17\% from the values given by %Arnett~(\cite{Ar91}).
%We considered the case of oxygen because the final abundance of this
%element is little affected by the stellar winds and rotation, two effects which are not 
%taken into account in the models by 
%Arnett~(\cite{Ar91}). 

%models with without rot
How does the contribution of the two models (rotating and non-rotating) at $Z=10^{-8}$ compare when both the wind
and the supernova contribute to the ejection of the stellar material?
First, one sees that the total mass ejected (through winds and supernova explosion)
is very similar (on the order of 54--55 M$_\odot$), 
due to the fact that the two models have similar CO core masses. Second,
one sees that
the amount of $^{4}$He ejected by the rotating model is slightly higher, whereas the amount of $^{12}$C is
lower due to the effect discussed above ($\alpha$-captures at the
end of the core He-burning phase). The quantity of $^{16}$O ejected is similar in both models.
Third, 
the most important differences between the rotating and 
non-rotating models
occur for $^{13}$C, $^{14}$N, $^{17}$O, and $^{18}$O. The abundances
of these isotopes are increased by factors between 10$^4$-10$^7$ in the ejecta of the rotating model.
The first three isotopes are
produced in the H-burning shell (CNO cycle) and are mainly ejected by the winds,
while the last one, produced at the interface between the CO-core and the He-burning shell, is ejected
at the time of the supernova explosion. 
Fourth, rotation also deeply affects the ratios of light elements in the ejected material (see Tables~\ref{tbl-2} \& \ref{tbl-3}).
The effects of rotation are qualitatively similar to those obtained when comparing the composition of the wind
material of rotating and non-rotating stellar models. Rotation decreases the 
$^{12}$C/$^{13}$C ratios from 3.5 $\times$ 10$^8$ in the non-rotating case to 311 in the rotating case,
while it increases the N/C and N/O ratios, which have values of $\sim$10$^{-7}$ and
$10^{-8}$  respectively, when $\upsilon_{\rm ini}$ = 0 km s$^{-1}$, and of 0.5 and 0.02 when $\upsilon_{\rm ini}$ = 800 km s$^{-1}$.

In the ejecta of rotating models,
composed of both wind and supernova material, the $^{12}$C/$^{13}$C ratio is higher than in pure wind material, and
the N/C and N/O ratios are smaller.
This comes from the fact that
the supernova ejecta 
are rich in helium-burning products characterized by a very high $^{12}$C/$^{13}$C ratio
and by very low N/C and N/O ratios.

%Here we discussed the composition of wind and supernova ejecta taken together. 
At this point we can ask if the rotating star
had lost no mass through stellar winds, and if all the stellar material were ejected
at the time of the supernova explosion,
would the composition of the ejecta be
different with respect to the case discussed above, where part of the material
is ejected by the winds and part by the supernova explosion. 
Let us recall that stellar winds remove layers from the stars at an earlier evolutionary stage than do
supernova explosions. If some of these layers, instead of being ejected by the winds, had
remained locked inside the star, they would have been processed further by the nuclear reactions. 
Thus their composition
at the end of the stellar evolution
would be different from the one obtained if they had been ejected at an earlier time by the winds.
Obviously for such differences to be important, mass loss must remove the layers
at a sufficiently early time. If it does so only at the very end of the evolution, there would be no
chance for the layers to be processed much by the nuclear reactions, and there would be no significant
difference whether the mass were ejected by the winds or by the supernova explosion.
Actually, this is what happens in our rotating models. As indicated above, the mass
is removed at the very end of the He-burning phase, and only material from
the H-rich envelope is ejected. Thus, if this material were ejected only
at the time of the supernova explosion, it would have kept the same chemical composition
as the one in Table~\ref{tbl-2}. 

As a result, the chemical composition of the ejecta (wind and supernova) does
not depend much on the mass loss, but is deeply affected by rotation.
However,
the stellar winds may of course be of primary importance 
if the whole final mass of the star is swallowed in a
black-hole at the end of the evolution. In that case,
the star will contribute to the interstellar enrichment only by
its winds.

\begin{table}
\caption{Helium-, CO-core mass and mass of the remnants 
(respectively $m_\alpha$, $m_{\rm CO}$, and $m_{\rm rem}$)
of 60 M$_\odot$ stellar models with
and without rotation at $Z=10^{-8}$.
The total mass ejected ($m_{\rm ej}$) and the mass ejected of various chemical species
($m(X_i)$)
are given in solar masses. The values of some isotope ratios (in mass fractions) are also indicated.
The case of matter ejected by stellar winds only is distinguished from the case
of matter ejected by both the
stellar winds and the supernova explosion.} \label{tbl-2}
\begin{center}\scriptsize
\begin{tabular}{|c|ll|ll|}
\hline
  &  &  &  &    \\
    &  \multicolumn{2}{|c|}{$M_{\rm ini}$/M$_\odot$ $\ $ $Z$  $\ \ \ $ $\upsilon_{\rm ini}$}
    &  \multicolumn{2}{|c|}{$M_{\rm ini}$/M$_\odot$ $\ $ $Z$  $\ \ \ $ $\upsilon_{\rm ini}$}
\\
    &  \multicolumn{2}{|r|}{$\left[{{\rm km}\over {\rm s}}\right]$}
    &  \multicolumn{2}{|r|}{$\left[{{\rm km}\over {\rm s}}\right]$}
\\
  &  &  &  &    \\
    &  \multicolumn{2}{|c|}{$\ \ \ \ \ $60$\ $  $\ \ $ 10$^{-8}$  $\ \ \ $   0 $\ $} 
    &  \multicolumn{2}{|c|}{$\ \ \ \ \ $60$\ $  $\ \ $ 10$^{-8}$  $\ $ 800 $\ $}
\\  
  &  &  &  &    \\  
\hline
\hline
  &  &  &  &    \\
$m_\alpha$    &\multicolumn{2}{|c|}{23.08} &  \multicolumn{2}{|c|}{23.83}   \\
$m_{\rm CO}$  &\multicolumn{2}{|c|}{21.61} &  \multicolumn{2}{|c|}{18.04}    \\  
$m_{\rm rem}$ &\multicolumn{2}{|c|}{6.65} &  \multicolumn{2}{|c|}{5.56}   \\  
  &  &  &  &    \\
\hline
\hline
  &  &  &  &    \\
 &\multicolumn{2}{|c|}{Mass ejected} &  \multicolumn{2}{|c|}{Mass ejected}   \\    
  &  &  &  &    \\
   &  WIND & SN+WIND
    &  WIND & SN+WIND 
\\
  &  &  &  &    \\ 
$m_{\rm ej}$& 0.28      & 53.35      & 36.17     &  54.44     \\    
%$^3$He         & 7.12e-06  & 6.80e-05   & 7.25e-06  &  7.25e-06  \\
$m(  ^4{\rm He})$            & 6.62e-02  & 19.58      & 21.46     &  23.85     \\
$m(^{12}{\rm C})$            & 2.08e-10  & 2.066      & 4.78e-03  &  4.26e-01  \\
$m(^{13}{\rm C})$            & 2.84e-12  & 5.84e-09   & 1.25e-03  &  1.37e-03  \\
$m(^{14}{\rm N})$            & 6.44e-11  & 1.90e-07   & 1.97e-01  &  2.20e-01  \\
$m(^{16}{\rm O})$            & 1.85e-09  & 12.61      & 6.08e-03  &  13.54     \\
$m(^{17}{\rm O})$            & 8.27e-13  & 6.39e-10   & 7.50e-06  &  8.60e-06  \\
$m(^{18}{\rm O})$            & 4.14e-12  & 6.53e-10   & 1.58e-08  &  4.44e-03  \\
  &  &  &  &    \\
\hline
\hline
  &  &  &  &    \\
 &\multicolumn{2}{|c|}{Isotopic ratios} &  \multicolumn{2}{|c|}{Isotopic ratios}   \\    
  &  &  &  &    \\
$^{12}$C/$^{13}$C   & 73.24     & 3.54e+08   & 3.82      &  311       \\
N/C                 & 0.31      & 9.20e-08   & 41.2      &  0.52      \\
N/O                 & 0.03      & 1.51e-08   & 32.4      &  0.02      \\
  &  &  &  &    \\  
\hline

\end{tabular}
\end{center}

\end{table}

\begin{table}
\caption{Same as Table~\ref{tbl-2} for rotating stellar models at $Z=10^{-5}$.
The two 60 M$_\odot$ models were computed with different physical ingredients, see text.
} \label{tbl-3}
\begin{center}\scriptsize
\begin{tabular}{|c|ll|ll|}
\hline
  &  &  &  &    \\
    &  \multicolumn{2}{|c|}{$M_{\rm ini}$/M$_\odot$ $\ $ $Z$  $\ \ \ $ $\upsilon_{\rm ini}$}
    &  \multicolumn{2}{|c|}{$M_{\rm ini}$/M$_\odot$ $\ $ $Z$  $\ \ \ $ $\upsilon_{\rm ini}$}
\\
    &  \multicolumn{2}{|r|}{$\left[{{\rm km}\over {\rm s}}\right]$}
    &  \multicolumn{2}{|r|}{$\left[{{\rm km}\over {\rm s}}\right]$}
\\
  &  &  &  &    \\
    &  \multicolumn{2}{|c|}{$\ \ \ \ \ $60$\ $  $\ \ $ 10$^{-5}$  $\ \ \ $ 800 $\ $} 
    &  \multicolumn{2}{|c|}{$\ \ \ \ \ $60$\ $  $\ \ $ 10$^{-5}$  $\ $ 800 $\ $}
\\  
  &  &  &  &    \\  
\hline
\hline
  &  &  &  &    \\
$m_\alpha$    &\multicolumn{2}{|c|}{36.90} &  \multicolumn{2}{|c|}{30.69}   \\
$m_{\rm CO}$  &\multicolumn{2}{|c|}{28.60} &  \multicolumn{2}{|c|}{27.95}    \\  
$m_{\rm rem}$ &\multicolumn{2}{|c|}{8.69} &  \multicolumn{2}{|c|}{8.50}   \\  
  &  &  &  &    \\
  \hline
  \hline
  &  &  &  &    \\ 
 &\multicolumn{2}{|c|}{Mass ejected} &  \multicolumn{2}{|c|}{Mass ejected}   \\    
  &  &  &  &    \\
   &  WIND & SN+WIND
    &  WIND & SN+WIND 
\\
  &  &  &  &    \\ 
$m_{\rm ej}$& 23.10     & 51.31      & 29.31     &  51.50     \\    
%$^3$He         & 4.59e-05  & 4.61e-05   & 8.88e-05  &  8.88e-05  \\
$m(^4{\rm He})$         & 12.20     & 18.99      & 12.60     &  14.58     \\
$m(^{12}{\rm C})$       & 3.34e-04  & 5.84e-01   & 1.45e-02  &  2.46      \\
$m(^{13}{\rm C})$       & 6.81e-05  & 1.47e-04   & 3.81e-03  &  2.58e-02  \\
$m(^{14}{\rm N})$       & 9.78e-03  & 2.51e-02   & 4.29e-02  &  1.87e-01  \\
$m(^{15}{\rm N})$       & 3.21e-07  & 2.15e-06   & 1.55e-06  &  1.68e-05  \\
$m(^{16}{\rm O})$       & 2.72e-04  & 18.12      & 3.29e-02  &  17.32     \\
$m(^{17}{\rm O})$       & 4.59e-07  & 2.40e-06   & 2.78e-05  &  1.32e-04  \\
$m(^{18}{\rm O})$       & 2.82e-08  & 1.34e-03   & 1.63e-08  &  2.07e-04  \\
$m(^{19}{\rm F})$       & 1.95e-09  &            & 1.10e-08  &        \\
$m(^{20}{\rm Ne})$      & 7.64e-06  &            & 1.29e-05  &        \\
$m(^{21}{\rm Ne})$      & 1.84e-08  &            & 6.99e-08  &        \\
$m(^{22}{\rm Ne})$      & 4.80e-07  &            & 3.35e-05  &        \\
$m(^{23}{\rm Na})$      & 1.22e-06  &            & 5.61e-06  &        \\
$m(^{24}{\rm Mg})$      & 4.41e-06  &            & 6.21e-06  &        \\
$m(^{25}{\rm Mg})$      & 3.95e-07  &            & 6.96e-07  &        \\
$m(^{26}{\rm Mg})$      & 5.68e-07  &            & 3.07e-06  &        \\
$m(^{27}{\rm Al})$      & 5.49e-06  &            & 7.75e-06  &        \\
  &  &  &  &    \\
\hline
\hline
  &  &  &  &    \\
 &\multicolumn{2}{|c|}{Isotopic ratios} &  \multicolumn{2}{|c|}{Isotopic ratios}   \\    
  &  &  &  &    \\  
$^{12}$C/$^{13}$C   & 4.90      & 3970       & 3.81      &  95.3      \\
 N/C                & 29.3      & 0.04       & 2.96      &  0.08      \\
 N/O                & 36.0      & 0.001      & 1.30      &  0.01      \\
&  &  &  &    \\  
\hline

\end{tabular}
\end{center}

\end{table}

Comparing the data given in the right part of Table~\ref{tbl-2} with the left part of Table~\ref{tbl-3},
one can see the effect produced by an increase in the initial metallicity by three orders
of magnitude (all other things being equal). 
Interestingly, we see that the differences between the two models are in general 
much smaller than those between the rotating
and non-rotating model at a given metallicity. The total ejected mass, and the masses of $^4$He, $^{12}$C,
$^{16}$O are similar within factors between 0.8 and 1.4. The quantities of $^{14}$N and $^{13}$C are
within an order of magnitude, and the masses of $^{17}$O and $^{18}$O differ by a factor 3.
We are thus far from the factors 10$^4$-10$^7$ between the results of the rotating and non-rotating models
at $Z=10^{-8}$ ! The effects of rotation at extremely low metallicity are much larger than the effects of a change in the initial $Z$ content.

The results given on the right side of Table~\ref{tbl-3} corresponds to the model
described in Meynet et al.~(\cite{Meynetal05}). It differs from
the present models by the mass loss and mixing prescription (see Sect.~3.2). As already emphasized above, the results are qualitatively very similar.
However, quantitatively, they present some differences. For instance, the quantity of $^{12}$C in the model
presented on the right side of Table~\ref{tbl-3} is larger by a factor 4
compared to the value given on the left side of the same Table.
The right model presents a smaller helium core,
an effect mainly due to higher mass loss rates. This favours larger ejections of carbon by the winds and also by the supernova, since
smaller helium cores lead to higher
C/O ratios at the end of the helium-burning phase.
In the right model, 
the quantity
of $^{16}$O is decreased by about 4\%. The ejected masses of $^{13}$C and $^{17}$O are increased by factors of 176 and 55, respectively.
The masses of the other isotopes 
differ by less than
an order of magnitude.

\section{Link with the extremely metal--poor C-rich stars}

\subsection{Observations and existing interpretations}

Spectroscopic surveys of very metal--poor stars (Beers et al.~\cite{Be92}; 
Beers~\cite{Be99}; Christlieb~\cite{Ch03}) 
have shown that CEMP stars
account for up to about 25\% of stars with metallicities lower than 
[Fe/H]$\sim -2.5$ (see e.g. Lucatello et al.~\cite{Lu04}).
A star is said to be C-rich if [C/Fe]$>1$.
A large proportion 
of these CEMP stars also present enhancements in their neutron capture elements
(mainly $s$-process elements). A few of them also appear to exhibit large 
enhancements in N and O. The most iron-deficient stars observed so far are CEMP stars.
These stars are 
HE 0107-5240, a giant halo star, and HE 1327-2326, a dwarf or subgiant halo star. 
The star HE 0107-5240 ([Fe/H]=-5.3) presents the following CNO surface abundances: [C/Fe]=4.0, [N/Fe]=2.3, and
[O/Fe]=2.4 (Christlieb et al.~\cite{christ04}; 
Bessell et al.~\cite{bessel04}). The ratio $^{12}$C/$^{13}$C 
has also been tentatively estimated by
Christlieb et al.~(\cite{christ04}), who suggest a value of about
60, but with a great uncertainty. They can, however, rule out a value inferior to 50
(let us recall that the solar ratio is $\sim$73).
The star HE 1327-2326 has [Fe/H]=-5.4 and CNO surface abundances: [C/Fe]=4.1, [N/Fe]=4.5,
[O/Fe]$< 4.0$ (Frebel et al.~\cite{Fr05}).

The origin of the high carbon abundance is still an open question and various scenarios have been
proposed: 
\begin{enumerate}
\item {\bf The primordial scenarios}: in this case the abundances observed at the surface
of CEMP stars are the abundances of the cloud from which the star formed. The protostellar
cloud was enriched in carbon by one or a few stars from a previous generation.
For instance, Umeda and Nomoto (\cite{Um03}) propose that the cloud from which HE 0107-5240
formed was enriched by the ejecta of one Pop III 25 M$_\odot$ star, which had exploded with low
explosion energy (on the order of 3 $\times 10^{50}$ erg) and experienced 
strong mixing and fallback at the time of the supernova explosion.
The mixing is necessary to create the observed 
high--level enrichments in light elements, and the fallback is necessary 
to retain a large part of the iron peak elements. 
Limongi et al (\cite{Li03}) suggest that the cloud was enriched by the ejecta of two supernovae
from progenitors
with masses of about 15 and 35 M$_\odot$. 
\item {\bf The accretion/mass transfer scenarios}: some authors have proposed that this particular 
abundance pattern results from accretion of 
interstellar material and from a companion (for instance an AGB star, as proposed by Suda et al.~\cite{Su04}).
As far as the nucleosynthetic origin is concerned, this scenario is not fundamentally different from the first one. 
\item {\bf The in situ scenarios}: finally, some authors have explored the possibility that
the star itself could have produced the particular abundance pattern
seen at its surface
(Picardi et al.~\cite{Pi04}). 
The overabundance of nitrogen might easily be explained in the frame of this scenario,
if the star had begun its evolution with the high carbon and oxygen overabundance. In fact we did perform a test calculation
of a non-rotating 0.8 M$_\odot$ stellar model at [Fe/H]=-5.3 with an initial value of [C/Fe] and [O/Fe] equal to
 4.0 and 2.3, respectively, {\it i.e.} equal to the abundances observed at the surface of HE 0107-5240. We found that, when
the star reaches the value of the effective temperature ($T_{\rm eff}$ = 5100$\pm$ 150 K) and
of gravity ($\log g= 2.2\pm 0.3$) of HE  0107-5240 (Christlieb et al.~\cite{Ch02}),
the surface nitrogen enrichment is well within the range of the observed values. 
However, it appears difficult to invoke similar processes to explain the high carbon and oxygen
enhancements (see Picardi et al.~\cite{Pi04}).
\end{enumerate}

\subsection{No ``in situ'' CN production}

An abundance pattern typical of CEMP stars has been
observed at the surface of non-evolved stars (Norris et al.~\cite{norr97}; Plez \& Cohen \cite{Pl05};
Frebel et al.~\cite{Fr05}). Among the most recent observations, let us mention
the subgiant or dwarf star HE 1327-2326 (Frebel et al.~\cite{Fr05}, see above) and the dwarf
star G77-61 (Plez \& Cohen \cite{Pl05}).
The initial mass of G77-61 is estimated to be between 0.3 and 0.5 M$_\odot$, and it has an [Fe/H]=-4.03,
[C/Fe]=2.6, [N/Fe]=2.6, and a $^{12}$C/$^{13}$C ratio of 5$\pm$1.
In this case, there is no way for the star, which burns
its hydrogen through the pp chains, to produce nitrogen. 
There is even less possibility of producing surface enhancements of carbon and oxygen. 
Therefore, the ``in situ'' scenario can be excluded, at least for this star.
In that case, only the first and second scenarios are possible. 
The same is true for explaining the very high overabundances of carbon
and nitrogen at the surface of HE 1327-2326.

The values observed at the surface of non-evolved and evolved
stars are shown in Fig.~\ref{vent}.
In the second case,
the surface may have been depleted in carbon and oxygen and enriched in nitrogen due
to the dredge-up that occurs along the red giant branch. On the other hand, in the case of
the non-evolved stars, as explained above,
this mechanism cannot be invoked, and the measured abundances reflect
the abundances of the cloud that gave birth to the star. On the whole, the distributions of elements
is similar for evolved and non-evolved stars, which favours the
primordial scenario. 

In the following, we explore the first two scenarios using our fast--rotating models.
The abundance pattern observed at the surface 
of CEMP stars seems to be a mixture of hydrogen and 
helium burning products. Since rotation
allows these products to 
coexist in the outer layers of stars 
(both in massive and intermediate mass stars), this seems a useful direction
for our research.

\subsection{Comparison with wind composition}

\begin{figure}
\resizebox{\hsize}{!}{\includegraphics[angle=-90]{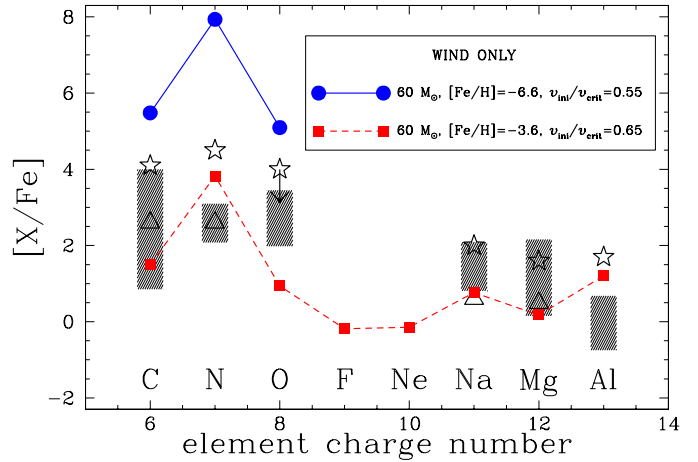}}
%\resizebox{\hsize}{!}{\includegraphics{ytotclair.jpg}}
\caption{Chemical composition of the wind
of rotating 60~M$_\odot$ models (solid circles and squares).
The hatched areas correspond to the range of values
measured at the surface of giant CEMP stars: HE 0107-5240, [Fe/H]$\simeq$~-5.3 
(Christlieb \& al. \cite{christ04});
CS 22949-037, [Fe/H]$\simeq$~-4.0 (Norris \& al.
\cite{norr01}; Depagne \& al. \cite{dep02}); CS 29498-043, [Fe/H]$\simeq$~-3.5 (Aoki
\& al. \cite{aoki04}). The empty triangles (Plez \& Cohen~\cite{Pl05}, [Fe/H]$\simeq -4.0$) 
and stars (Frebel et al.~\cite{Fr05},  [Fe/H]$\simeq -5.4$, only an upper limit is given for [O/Fe]) correspond to
non-evolved CEMP stars (see text).}
\label{vent}
\end{figure}

Let us first see if the CEMP stars could be formed from material made up of massive star winds,
or at least heavily enriched by winds of massive stars. At first sight, such
a model might appear quite unrealistic, since the period of strong stellar winds
is rapidly followed by the supernova ejection, which would add the ejecta of the supernova itself
to the winds. However,
for massive stars at the end of their nuclear lifetime, a black
hole, that swallows
the whole final mass, might be produced. In that case, the massive star
would contribute to the local chemical enrichment of the interstellar medium only through
its winds. Let us suppose that such a situation has occurred and that
the small halo star we observe today was
formed from the shock induced by the stellar winds with the interstellar material.
What would its chemical composition be? 
Its iron content would be the same as the iron abundance of
the massive star. Indeed,
the iron abundance of the interstellar medium would have no
time to change much in the brief massive star lifetime, 
and the massive star wind ejecta 
are neither depleted nor enriched in iron.
 
%***********************
The abundances of the other elements in the stellar winds for our two rotating 60 M$_\odot$
at $Z=10^{-8}$ and 10$^{-5}$ are shown in Fig.~\ref{vent}.
The ordinate [X/Fe] is given by the following expression:
$$[{\rm X/Fe}]=\log\left({{\rm X} \over {\rm X}_\odot}\right)-\log\left({X({\rm Fe}) \over X({\rm Fe})_\odot}\right),$$
where X is the mass fraction of element X 
in the wind ejecta, and X$_\odot$ in the Sun.
Similarly, the symbols $X$(Fe) and $X$(Fe)$_\odot$ refer to the mass fraction
of $^{56}$Fe
in the wind material or in the Sun.
Here we suppose that $\log(X({\rm Fe})/X({\rm Fe})_\odot)\sim[{\rm Fe/H}]$, since the mass fraction of hydrogen 
remains approximately constant, whatever the metallicity between $Z=$10$^{-8}$ and 0.02.
The values of [Fe/H] are those corresponding to the initial metallicity of the models
(for $Z$ = 10$^{-8}$ one has $\log(X({\rm Fe})/X({\rm H}))=-9.38$, see Table~\ref{tbl-0}).
The solar abundances are those chosen by Christlieb et al (\cite{christ04}) and Bessell et al. (\cite{bessel04})
in their analysis of the star HE 0107-5240, and they correspond to the solar abundances obtained recently by Asplund et al.~(\cite{AS05}).
In particular, $\log(X({\rm Fe})/X({\rm H}))_\odot=-2.80$, thus [Fe/H]=-6.6 at $Z$ = 10$^{-8}$, and
[Fe/H]=-3.6 at $Z$ = 10$^{-5}$.

From Fig.~\ref{vent}, we see that the winds are strongly enriched in CNO elements. 
The model at $Z=10^{-5}$, computed with an extended nuclear 
reaction network, allows us to look at the abundances in the winds of heavier elements. 
The wind material is also somewhat enriched 
in Na and Al. Before comparing with the observations, let us first note:
\begin{itemize}
\item 1) The more metal--poor model is shifted toward higher values compared 
to the metal rich one. If, in both models, the mass fraction of element X were
the same in the wind ejecta, then one would expect a shift by 3 dex when the iron content of the
model goes from [Fe/H]=-3.6 to [Fe/H]=-6.6. The actual shift is 
approximately 3.6 dex, slightly more than the
iron content difference of 3 dex between the two models. 
The additional 0.6 dex comes from the more efficient mixing
in the metal--poorer models. 
\item 2) In the frame of the hypotheses made here, {\it i.e.} a halo
star made from the wind ejecta that triggered its formation, we should compare
the wind composition from a massive star model with the same initial iron content as in the halo
star we considered. The range of iron contents in the models, [Fe/H] equal from -6.6 to -3.6,
covers the range
of iron content of the CEMP plotted in Fig.~\ref{vent}, whose [Fe/H] are between
-5.4 and -3.5. However, the [Fe/H] = -6.6 model is well below the lower bound
of the observed [Fe/H], making this model less interesting for comparisons
with the presently available observations. In that respect the [Fe/H] = -3.6 model, which has
an iron content that is comparable to the iron--richest stars observed, is more interesting.
\item 3) Any dilution with some amount
of interstellar material would lower the abundance of the element X without changing the mass
fraction of iron. In that case the values plotted in Fig.~\ref{vent} are shifted to lower values, but
the relative abundances of the elements will not change
(as far as the main source of the elements considered
are the wind ejecta). 
\end{itemize}
Keeping in mind these three comments, it appears that what has to be compared with
the observations are more the relative abundances between the CNO elements than the actual
values of the [X/Fe] ratios, which will depend on the initial metallicity of the model
considered, as well as on the dilution factor.

%***********************
%We see that for carbon and oxygen, the predicted wind composition 
%for the two metallicities well frame the abundances observed at the surface
%of CEMP stars. For nitrogen, the theoretical values 
%are above the range of values for the giant halo stars considered here
%(see caption of Fig.~\ref{vent}), but well frame the value observed at the surface
%of the dwarf or subgiant star HE 1327-2326.
%The model at $Z=10^{-5}$, computed with an extended nuclear 
%reaction network, allows us to look at the abundances in the winds of heavier isotopes. 
%We see that wind material is enriched 
%in Na and Al: for sodium the theoretical value is below the lower 
%observed limit, while for aluminium the theoretical value
%is at the upper limit of the observed values for giant stars, but below the value
%observed at the surface of HE 1327-2326. 
%

From Fig.~\ref{vent}, and Tables~\ref{tbl-2} \& \ref{tbl-3}, one sees that,
for the two metallicities considered here,
the wind material of rotating models is characterised by N/C and N/O ratios between
$\sim$ 1 and 40 and by $^{12}$C/$^{13}$C ratios around 4-5. These values are
compatible with the ratios observed at
the surface of CS 22949-037 (Depagne et al.~\cite{dep02}):
N/C $\sim$ 3 and $^{12}$~C~/~$^{13}$~C $\sim$4. The observed value
for N/O ($\sim$0.2) is smaller than the range of theoretical values, but greater
than the solar ratio ($\sim$ 0.03). Thus the observed N/O ratio
also bears the mark of some CNO processing, although slightly less
pronounced than in our stellar wind models. 

On the whole, 
a stellar wind origin for the material composing this star
does not appear out of order in view of these comparisons,
all the more so
if one considers that, in the present comparison, there is no
fine tuning of some parameters in order to obtain the best agreement possible. 
The theoretical results are
directly compared to the observations. Moreover only a small subset of possible initial conditions
has been explored.

Other CEMP stars present, however, lower values for the N/C and N/O ratios
and higher values for the $^{12}$C/$^{13}$C ratio.
For these cases, it appears that the winds of our rotating 60 M$_\odot$ models
appear to be too strongly CNO--processed (too
high N/C and N/O ratios and too low $^{12}$C/$^{13}$C ratios).
Better agreement would
be obtained if the observed abundances also stem from material coming from the CO-core and ejected
either by strong late stellar winds or in a supernova explosion.

\subsection{Expression of abundance ratios in total ejecta (winds and supernova)}

To find the initial chemical composition of stars that
would form from a mixture of wind and supernova ejecta 
with interstellar medium material,
let us define ${X}_{\rm ej}$ as the mass fraction of element X in the ejecta 
(wind and supernova).
This quantity can be obtained from the stellar models and computed according to the expression below:
$${\it X}_{\rm ej}={{\it X}_{\rm wind}  m_{\rm wind}+{\it X}_{\rm SN} m_{\rm SN} \over m_{\rm wind}+m_{\rm SN}},$$
where $X_{\rm wind}$ and $X_{\rm SN}$ are the mass fractions of element X in the wind,
in the supernova ejecta. Here $m_{\rm wind}$ and $m_{\rm SN}$ are the masses ejected by the stellar winds
and at the time of the supernova explosion.
To obtain the mass of the remnants, we adopted the relation obtained by Arnett~(\cite{Ar91}) between the masses
of the remnant and the CO core. This method is the same as the 
one adopted by Maeder (\cite{Ma92}). 

The total mass ejected by the star, $m_{\rm ej}=m_{\rm wind}+m_{\rm SN}$, is mixed with some
amount of interstellar material
$m_{\rm ISM}$. The mass fraction of element X in the material composed from the ejecta mixed with the interstellar medium will
be
$$X={X_{\rm ej} m_{\rm ej} + X_{\rm ini} m_{\rm ISM} \over m_{\rm ej} +m_{\rm ISM}}={X_{\rm ej}{m_{\rm ej} \over m_{\rm ISM}}+X_{\rm ini}\over
{m_{\rm ej} \over m_{\rm ISM}} +1},$$
where $X_{\rm ini}$ is the mass fraction of element X in the interstellar medium. 
In our case the interstellar medium is very metal poor
so that one can consider 
$X_{\rm ini}\sim 0$ for the heavy elements synthesised in great quantities by the star (note that this cannot be done for nitrogen ejected by the non-rotating
60 M$_\odot$ stellar model).
We also suppose that $m_{\rm ej}\ll m_{\rm ISM}$ and thus $X= (X_{\rm ej}m_{\rm ej})/m_{\rm ISM}$.
Using these expressions, one can write
$$[{\rm Fe/H}]=\log\left({X({\rm Fe})_{\rm ej}\over X({\rm Fe})_\odot}\right)+\log\left({m_{\rm ej}\over m_{\rm ISM}}\right),$$
assuming, as we did above, that $X({\rm H})_\odot /X({\rm H})_{\rm ej}\approx 1$.
Values of [X/Fe] are obtained using the expression
$$[{\rm X/Fe}]=[{\rm X/H}]-[{\rm Fe/H}]=$$
$$\log\left({{X}_{\rm ej}\over {X}_\odot}\right)-\log\left({X({\rm Fe})_{\rm ej}\over X({\rm Fe})_\odot}\right).$$
One needs to have an estimate for both $m_{\rm ej}/m_{\rm ISM}$ (the dilution factor)
and for the mass fraction of iron in the ejecta $X$(Fe)$_{\rm ej}$. 
A precise quantitative determination of $X$(Fe)$_{\rm ej}$ and $m_{\rm ej}/m_{\rm ISM}$ from
theory is quite difficult. For instance,
for a given initial mass, the quantity of iron ejected by the supernova
can vary by orders of magnitudes depending on the mass cut, the energetics of the supernova, and the geometry
of the explosion (see e.g. Maeda \& Nomoto \cite{MN03}). On the other hand, the dilution factor will depend
 on the energetics of the supernova, among other parameters. 
In the absence of
any very precise guidelines, we determined the two unknown quantities, 
the dilution factor, and the mass of ejected iron, requiring that the mixture will have
[Fe/H]=~-5.4 and [O/Fe]=~+3.5. The first value 
corresponds to the values observed at the surface
of the star HE 1327-2326 (Frebel et al.~\cite{Fr05}), and the second value
is below the upper limit of [O/Fe] ($< 4.0$) found for this star.
Doing so, one can write,
$$[{\rm X/Fe}]=[{\rm X/O}]+[{\rm O/Fe}]=$$
$$\log\left({{X}_{\rm ej}\over {\it X}_\odot}\right)-\log\left( {X({\rm O})_{\rm ej} \over X({\rm O})_\odot}\right) +3.5,$$
where $X$(O) is the mass fraction of oxygen. 
The mass fraction of ejected iron can be estimated
from
$$[{\rm O/Fe}]=
\log\left({X({\rm O})_{\rm ej}\over X({\rm O})_\odot}\right)-\log\left({X({\rm Fe})_{\rm ej}\over X({\rm Fe})_\odot}\right)=3.5,$$
and the dilution factor can be obtained from
$$[{\rm Fe/H}]=\log\left({X({\rm Fe})_{\rm ej}\over X({\rm Fe})_\odot}\right)+\log\left({m_{\rm ej}\over m_{\rm ISM}}\right)=-5.4.$$

\subsection{Results from the ``wind plus supernova ejecta'' model}

 \begin{figure}
\resizebox{\hsize}{!}{\includegraphics[angle=-90]{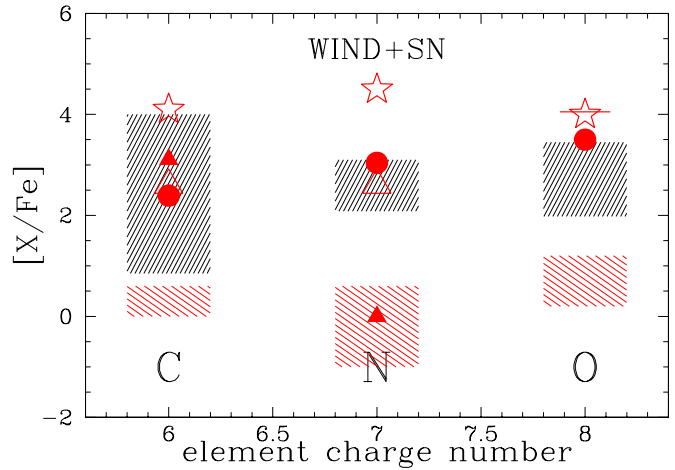}}
%\resizebox{\hsize}{!}{\includegraphics{ytotclair.jpg}}
\caption{Chemical composition of the ejecta
(wind and supernova) 
of 60~M$_\odot$ models: solid circles and triangles
correspond to models at $Z=10^{-8}$ ([Fe/H]=-6.6) with and without rotation.
The [N/Fe] ratio for the non-rotating model is equal to 0,
{\it i.e.}, no N-enrichment is expected.
The hatched areas (moving from top right down to the left) correspond to the range of values
measured at the surface of giant CEMP stars (same stars as in Fig.~\ref{vent}).
The empty triangles (Plez \& Cohen~\cite{Pl05}) 
and stars (Frebel et al.~\cite{Fr05}, only an upper limit is given for [O/Fe]) correspond to
non-evolved CEMP stars (see text).
The hatched areas (L to R from top) show the range of values
measured at the surface of normal halo giant stars by Cayrel et al.~(\cite{cayr04})
and Spite et al.~(\cite{Sp05}, unmixed sample only, see text).
}
\label{sn}
\end{figure}

Using the above formulae, let us now discuss what can be expected for the chemical composition
of a very metal--poor star formed from such a mixture.
This includes the CNO elements
for which the present models can give consistent estimates and the case of our models
at $Z=10^{-8}$ ([Fe/H]=-6.6), which are the models that are compatible with the requirement
that the mixture of ejecta and ISM material must have an [Fe/H]=~-5.4. Obviously
our second series of models at $Z=10^{-5}$ ([Fe/H]=~-3.6) does not fit
this requirement. Imposing [O/Fe]=3.5 and [Fe/H]=~-5.4 implies masses
of iron that are being ejected on the order of 1 $\times$ 10$^{-3}$ M$_\odot$ and mixed
with a mass of interstellar medium of about 2 $\times$ 10$^{5}$ M$_\odot$.
The mass of ejected iron (actually in the form of $^{56}$Ni) is very small
compared to the classical values of 0.07-0.10 M$_\odot$. On the other hand,
this quantity can be very small if a large part of the mass
falls back onto the
remnant (Umeda \& Nomoto~\cite{Um03}).
The mass of interstellar gas collected by the shock
wave can be related to the explosion energy $E_{\rm exp}$ through
(see Shigeyama \& Tsujimoto~\cite{Sh98})
$$M_{\rm ISM}=5.1 \times 10^4 {\rm M}_\odot \left({E_{\rm exp} \over 10^{51}{\rm erg}}\right).$$
%This expression assumes a sound speed of 10 km s$^{-1]$ (or a temperature of 10 000 K)
%for the interstellar gas .
A mass of 2 $\times$ 10$^{5}$ M$_\odot$ would correspond to
an energy equal to 4 $\times$ 10$^{51}$ erg, {\it i.e.}, a value well in the range
of energies released by supernova explosion. Thus imposing [O/Fe]=3.5 and [Fe/H]=-5.4
does not imply unrealistic values for the mass of iron that is ejected and for the mass of
interstellar medium swept up by the shock wave of the supernova explosion.

The theoretical ratios for the CNO elements are shown in Fig.~\ref{sn} and compared
with the ratios observed at the surface of CEMP stars and of normal giant halo stars
by Cayrel et al.~(\cite{cayr04}) and Spite et al.~(\cite{Sp05}). 
For oxygen, the value of 3.5
is obtained by construction, so it does not provide any constraint; however, see the previous
paragraph.
More interesting, of course, are the carbon and nitrogen abundances.
One sees that both the non-rotating and rotating models might account for some level of C-enrichment
that is compatible with the range of values observed at the surface of CEMP stars. However,
only the rotating models produce N-rich material at this level. 
Figure~\ref{sn} also shows that the predicted value for the N/C and N/O ratios 
from the rotating model appear to agree with the observed values
of these ratios at the surface of CEMP stars.
Thus as expected, the addition of 
material from the CO core (here ejected at the time of the supernova explosion)  to the wind material 
(mainly enriched in CNO--processed material), reduces
the N/C and N/O ratios.

The theoretical values in the wind--plus--supernova ejecta model for the ratio $^{12}$C/$^{13}$C are between 100 and 4000
for the rotating models (see Tables~\ref{tbl-2}
\& \ref{tbl-3}). 
The value predicted by the non-rotating model is much higher, on the order of 
10$^8$. Compared to the observed values, which are between 4 and 60, the value of the non-rotating model
is in excess by at least seven orders of magnitude. The situation is much more favourable for the rotating models.
In this last case,
the predicted values are still somewhat too high, but by much lower factors. 
 
Other proportions probably exist 
between wind and supernova mass ejecta than those considered here, 
which would provide a better fit to the observed surface abundances of CEMP stars.
Also more models need to be calculated for
exploring the set of initial parameters leading to a good 
correspondence between theory and observations.
From the large range of results obtained by different initial
conditions, there is little doubt,
that such a given set of parameters exists. 
The $^{12}$C/$^{13}$C ratio appears extremely sensitive to input parameters, so it may be
a powerful tool for a closer identification of the exact nucleosynthetic site.
%Would the mass of the remnant be more massive than what we have assumed here or the fall back be more important, then
%one would have obtained higher values for [N/Fe] and [C/Fe] and lower values for $^{13}$C/$^{12}$C as can be seen
%from Fig.~\ref{fallb}.

As can be seen from Fig.~\ref{sn}, the abundances observed at the surface of the normal giant stars by 
Cayrel et al.~(\cite{cayr04}) and Spite et al.~(\cite{Sp05}) are not far from solar ratios, and
are well below the range of values
observed at the surface of CEMP stars. Only the subset of stars qualified as unmixed by
Spite et al.~(\cite{Sp05}), {\it i.e.}, presenting no evidence of C to N conversion, has been
plotted here.
Probably these stars are formed from a reservoir of matter made up of the ejecta of
different initial mass stars, convolved with a proper distribution of the initial rotation velocities,
while the C-rich stars require some special circumstances involving a few or maybe only
one nucleosynthetic event.

\begin{figure}
\resizebox{\hsize}{!}{\includegraphics[angle=-90]{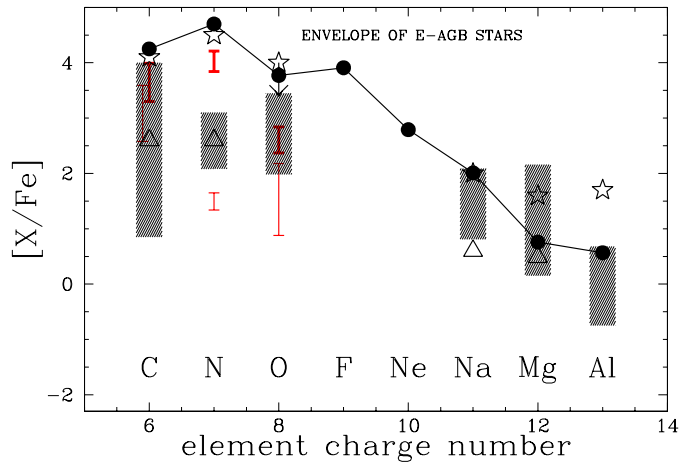}}
%\resizebox{\hsize}{!}{\includegraphics{ytotclair.jpg}}
\caption{Chemical composition of the envelopes of E-AGB stars compared to abundances
observed at the surface of CEMP stars (hatched areas). The continuous line shows the case
of a 7 M$_\odot$
at $Z=10^{-5}$ ([Fe/H]=-3.6) with $\upsilon_{\rm ini}=$ 800 km s$^{-1}$.
The vertical lines (shown as ``error bars'')
indicate the 
ranges of values for CNO elements
in the stellar models of Meynet \& Maeder~(\cite{MMVIII})
(models with initial masses between 2 and 7 M$_\odot$ at $Z=10^{-5}$).
The thick and thin lines correspond to rotating ($\upsilon_{\rm ini}$ = 300 km s$^{-1}$)
and non-rotating models.
The empty triangles (Plez \& Cohen~\cite{Pl05}) 
and stars (Frebel et al.~\cite{Fr05}, only an upper limit is given for [O/Fe]) correspond to
non-evolved CEMP stars.}
\label{agb}
\end{figure}

\subsection{Chemical composition of the envelopes of E-AGB stars}

One can wonder whether intermediate mass stars could also play a role
in explaining the peculiar abundance pattern of CEMP stars.
For instance, Suda et al.~(\cite{Su04}) propose that the small halo star,
observed today as a CEMP star, was the secondary in a close binary system. 
The secondary might have accreted matter from its evolved companion, an AGB star,
and might have thus acquired at least part of its peculiar surface abundance pattern.

The physical
conditions encountered in the advanced phases of an intermediate mass star
are not so different from the one realised in massive stars. Thus the same
nuclear reaction chains can occur and lead to similar nucleosynthetic products.
Also the lifetimes of massive stars (on the order of a few million years) are
not very different from the lifetimes of the most massive intermediate mass stars;
typically a 7 M$_\odot$ has a lifetime on the order of 40 Myr, only an order of magnitude higher
than a 60 M$_\odot$ star.
Moreover, the observation of s-process element overabundances at the surface of some
CEMP stars also point toward a possible Asymptotic Giant Branch (AGB) star origin\footnote{Note that
massive stars also produce s-process elements. The massive star s-process elements
have low atomic mass number (A inferior to about 90) and
are known under the name of the weak component of the s-process.}
for the material composing the CEMP stars.

To explore this scenario, we computed a
7 M$_\odot$ with $\upsilon_{\rm ini}=800$ km s$^{-1}$ at $Z=10^{-5}$, and
with the same physical ingredients as the 60 M$_\odot$ stellar models of the
present paper. In contrast to the 60 M$_\odot$ models, the 7 M$_\odot$ stellar model 
loses little mass during the core H- and He--burning phase, so that 
the star has still nearly its whole
original mass
at the
early asymptotic giant branch stage (the actual mass at this stage is 6.988 M$_\odot$). This is because
the star never reaches the break-up limit during the MS phase; and, 
due to rotational mixing and dredge-up,
the metallicity
enhancement at the surface  
only occurs very late, when the star
evolves toward the red part of the HR diagram
after the end
of the core He-burning phase. At this stage, the outer
envelope of the star is enriched in primary CNO elements, and the surface metallicity
reaches about 1000 times the initial metallicity.
If such a star is in a close
binary system, there is good chance that mass transfer occurs during this
phase of expansion of the outer layers. In that case, the secondary may accrete
part of the envelope of the E-AGB star. 

From the 7 M$_\odot$ stellar model, we can estimate the chemical composition 
of the envelope 
at the beginning of the thermal pulse AGB phase. Here we call
envelope all the material above the CO-core.
The result is shown in Fig.~\ref{agb} (continuous line with solid circles).
We also plotted the values obtained from the models
of Meynet \& Maeder~(\cite{MMVIII}) for 
initial masses between 2 and 7 M$_\odot$ at $Z=10^{-5}$
and with $\upsilon_{\rm ini}=0$ and $300$ km s$^{-1}$.

Before discussing the comparisons with observations, let us make two remarks:
1) as was the case for the theoretical predictions of the
massive star winds, the values given here have not been adjusted to
fit the observed values but result from the basic physics of the models; 2) the initial metallicity of our AGB models 
([Fe/H]=-3.6) is at the higher range of values of 
the metallicities observed for the CEMP stars. However,
based on the results from our massive star models at [Fe/H]=-6.6 and -3.6
(see Fig.~\ref{vent}), we see that
the overall pattern of the abundances will probably remain quite similar
for a lower initial metallicity. Only a shift toward higher values along the vertical axis
is expected when the initial metallicity of the model is decreased.

Looking at Fig.~\ref{agb}, one can note the three following points: 
\begin{enumerate}
\item The envelope of rotating intermediate mass stars presents a chemical composition in carbon, 
nitrogen, and oxygen that agrees well with the observed value at the surface of CEMP stars. In particular,
compared to the wind material of massive stars (see Fig.~\ref{vent}), the N/C ratios and N/O ratios
are in better agreement. The non-rotating models cannot account
for the high overabundances in nitrogen and oxygen.
\item The $^{12}$C/$^{13}$C ratios in our rotating models are between 19 and 2500, with the lowest values
corresponding to the most massive intermediate--mass star models. The non--rotating
models give values between 3 $\times$ 10$^5$ and 2 $\times$ 10$^6$.
Again here, rotating models agree much better with the observed values, although
very low $^{12}$C/$^{13}$C values (on the order of 4-5, as observed {\it e.g.} at the surface
of the dwarf halo star G77 61, see Plez and Cohen~\cite{Pl05}) 
seem to be reproduced only by massive star models (wind material only).
\item For sodium and aluminum, the ratios predicted by our 7 M$_\odot$ model
with $\upsilon_{\rm ini}=800$ km s$^{-1}$ fit the observed values well. In the case
of magnesium, good agreement is also obtained.
\end{enumerate}

Thus we see that the envelopes of AGB stellar models with rotation 
show a very similar chemical composition to the one observed at the surface
of CEMP stars. It is, however, still difficult to say that rotating intermediate mass star models
are better than rotating massive star models in this respect. Probably,
some CEMP stars are formed from massive star ejecta and others
from AGB star envelopes. Interestingly at this stage,
some possible ways to distinguish between massive star wind material
and AGB envelopes do appear. Indeed, we just saw above that
massive star wind material is characterised by a very low $^{12}$C/$^{13}$C ratio,
while intermediate mass stars seem to present higher values for this ratio.
The AGB envelopes would also present very high overabundances
of $^{17}$O, $^{18}$O, $^{19}$F, and $^{22}$Ne, while wind of massive rotating
stars present a weaker overabundance of $^{17}$O and depletion of 
$^{18}$O, $^{19}$F, and $^{22}$Ne.
As discussed in Frebel et al.~(\cite{Fr05}), the ratio of heavy elements, such
as the strontium--to--barium ratio, can also give clues to the origin of the material
from which the star formed. In the case of HE 1327-2326, Frebel et al.~(\cite{Fr05})
give a lower limit of [Sr/Ba] $> -0.4$, which suggests that strontium was not
produced in the main s-process occurring in AGB stars, thus leaving
the massive star hypothesis as the best option, in agreement with the result
from $^{12}$C/$^{13}$C in G77-61 (Plez \& Cohen~\cite{Pl05}) and
CS 22949-037 (Depagne et al.~\cite{dep02}).

\section{Conclusion}

We have proposed a new scenario for the evolution of very metal--poor massive stars.
This scenario requires no new physical processes, as it is based on models that have been 
extensively compared to observations of stars at solar composition and in the
LMC and SMC.
The changes with respect to classical scenarios are twofold and are both induced by fast rotation:
first, rotational mixing deeply affects the 
chemical composition of the material ejected by the massive stars;
second, rotation significantly enhances the mass lost by stellar winds.
The mass loss rates are increased mainly because the mixing process is so strong that the surface metallicity is
enhanced by several orders of magnitude. This leads to strong radiative winds during the
evolution in the red part of the HR diagram. The strongest mass loss occurs at
the very end of the core He-burning phase.
The proposed scenario may 
allow very massive stars
to avoid the pair instability. 

We show that material ejected
by rotating models has chemical compositions that show
close similarities to the peculiar 
abundance pattern observed at the surface of CEMP stars. 
We explored the three possibilities of
CEMP stars made of: 1) massive star wind material, 2) total massive star ejecta
(wind plus supernova ejecta), 3) and material from E-AGB star envelopes.
Interestingly, from the models computed here, one can order these
three possibilities according to the degree of richness in CNO processed material.
From the richest to the poorest, one has the wind material, the E-AGB envelope, and
the total ejecta of massive stars. The imprints on the abundance pattern of CEMP stars
are thus not the same, depending on which source is involved. There is good hope that
in the future, it will be possible to distinguish them.

Other interesting questions will be explored in the 
future with these rotating metal--poor models.
Among them let us briefly mention:
\begin{itemize}

\item {\it What is the enrichment in new synthesized helium by the first stellar generations?}
This is a fundamental question already asked long ago by
Hoyle \& Tayler (\cite{Ho64}).
A precise knowledge of the helium enrichment caused by the first massive stars
(Carr et al.~\cite{Ca84}; Marigo et al. \cite{Ma03})
is important in order to correctly deduce the value of the cosmological helium
from the observed abundance
of helium in low metallicity regions  
(see e.g. Salvaterra \& Ferrara \cite{Sa03}).
Production of helium by the first massive stars may also affect the initial helium content of stars in
globular clusters. If the initial helium content of stars in globular 
clusters is increased by 0.02 in mass fraction, the 
stellar models will provide ages for the globular 
clusters that are lower by roughly 15\%, {\it i.e.}, 2 Gyr  starting from
an age of 13 Gyr (Shi~\cite{Sh95}; see also the interesting discussion in Marigo et al.~\cite{Ma03}). 
In the case of our rotating 60 M$_\odot$ at $Z=10^{-8}$, 22\% of the initial mass is ejected in the form 
of new synthesised helium by stellar winds. Thus the models presented here will certainly lead to new views on the question of
the helium enrichment at very low metallicity, provided of course, they are representative of the evolution of
the majority of massive stars at very low $Z$.

\item{\it What are the sources of primary nitrogen in very metal--poor halo stars?}
Primary $^{14}$N
is produced in large quantities in our rotating models.
In our previous work on this subject (Meynet \& Maeder~\cite{MMVIII}), we discussed the yields from stellar models at $Z=10^{-5}$ with
$\upsilon_{\rm ini}=300$ km s$^{-1}$. Such an initial velocity corresponds to a ratio 
$\upsilon_{\rm ini}/\upsilon_{\rm crit}$ of only 0.25. This value is lower than the value of $\sim$0.35
reached by solar metallicity models with $\upsilon_{\rm ini}=300$ km s$^{-1}$.
With such a low initial ratio of $\upsilon_{\rm ini}/\upsilon_{\rm crit}$, we found that
the main sources of primary nitrogen were intermediate mass stars with initial masses around about 3 M$_\odot$. 
However, as already shown in Meynet \& Maeder (\cite{MMVIII}),
the yield in $^{14}$N increases rapidly when the initial velocity increases.
As a numerical example, the yield
in primary nitrogen for the $Z=10^{-5}$, 60 M$_\odot$ with $\upsilon_{\rm ini}/\upsilon_{\rm crit}$ equal to 0.25 
was 7 $\times 10^{-4}$ M$_\odot$, while
the corresponding model with $\upsilon_{\rm ini}/\upsilon_{\rm crit}$ equal to 0.65 produces a yield of $\sim$2 $\times 10^{-1}$; 
{\it i.e.}, it increased by a factor of nearly 300~! 
Interestingly, these high yields of primary $^{14}$N from short-lived massive stars
seem to be required for explaining the high N/O ratio
observed in metal--poor halo stars (Chiappini et al.~\cite{Ch05}). Note 
that massive intermediate mass stars, whose lifetimes would be only an order of magnitude higher
than those of the most massive stars, could also be invoked to explain the high N/O ratio observed in very 
metal--poor stars. The age-metallicity relation is not precise enough to allow us to distinguish between the two.

\item{\it How does rotation affect the yields of extremely metal--poor stars?}
Other elements such as $^{13}$C, $^{17}$O, $^{18}$O, and $^{22}$Ne 
are also produced in much greater quantities in the rotating models.
Computations are now in progress for extending the range of initial parameters explored and to study
the impact of such models on the production of these isotopes, as well as on 
s-process elements. 
\end{itemize}

We think that the fact that stars rotate, and may even rotate fast
especially at low metallicity, 
has to be taken into account for obtaining more realistic models of 
the extremely metal--poor stars that formed in the early life of the Universe.

\begin{acknowledgements}
The authors are grateful to Dr. Joli Adams for the careful
language editing of the manuscript.
\end{acknowledgements}

\end{document}